\documentclass[journal,draftclsnofoot,onecolumn,12pt]{IEEEtran}
\usepackage{amsthm,amssymb,graphicx,multirow,amsmath,color,amsfonts}
\usepackage[update,prepend]{epstopdf}
\usepackage[noadjust]{cite}
\usepackage[latin1]{inputenc}
\usepackage{tikz}
\usepackage{bbm} 
\usepackage{pdfpages}
\usepackage{multirow}
\usepackage{subfig}
\usepackage{comment}
\usepackage{amsmath}
\usepackage{nccmath}
\usepackage{hhline}
\def\nb0{{\mathbf{0}}}
\def\nb1{{\mathbf{1}}}

\newtheorem{lemma}{Lemma}

\newtheorem{nrem}{Remark}
\newtheorem{theorem}{Theorem}

\newtheorem{cor}{Corollary}

\allowdisplaybreaks 

\usepackage{setspace}	

\begin{document}
\graphicspath{{./Figures/}}
\title{
Coverage analysis of Tethered {UAV}-assisted Large Scale Cellular Networks
}
\author{
Safa Khemiri, Mustafa A. Kishk, {\em Member, IEEE} and Mohamed-Slim Alouini, {\em Fellow, IEEE}
\thanks{Safa Khemiri and Mohamed-Slim Alouini are with King Abdullah University of Science and Technology (KAUST), CEMSE division, Thuwal 23955-6900, Saudi Arabia (e-mail: safa.khemiri@kaust.edu.sa; slim.alouini@kaust.edu.sa).  Mustafa Kishk is with the Department of Electronic Engineering, National University of Ireland, Maynooth, W23 F2H6, Ireland. (e-mail: mustafa.kishk@mu.ie). This work was funded in part by the Center of Excellence for NEOM Research at KAUST.
} 
\vspace{-4mm}}

\maketitle

\begin{abstract}
One of the major challenges slowing down the use of  unmanned aerial vehicles (UAVs) as aerial base stations (ABSs) is the limited on-board power supply which reduces the UAV's flight time. Using a tether to provide UAVs with power can be considered a reasonable compromise that will enhance the flight time while limiting the UAV's mobility. In this work,  we propose a system where ABSs are deployed at the centers of user hotspots to offload the traffic and assist terrestrial base stations (TBSs). 
Firstly, given the location of the ground station in the user hotspot (user cluster) and the users spatial distribution, we compute the optimal inclination angle and length of the tether. Using these results, we compute the densities of the tethered {UAV}s deployed at different altitudes, which enables tractable analysis of the interference in the considered setup.  Next, using tools from stochastic geometry and an approach of dividing user clusters into finite frames, we analyze the coverage probability as a function of the maximum tether length, the density of accessible rooftops for {UAV} ground station deployment, and the density of clusters. We verify our findings using Monte-Carlo simulations and draw multiple useful insights. For instance, we show that it is actually better to deploy {UAV}s at a fraction of the clusters, not all of them as it is usually assumed in literature.
\end{abstract}

\section{Introduction}
\bigskip
\bigskip
The use of unmanned aerial vehicles (UAVs) in a variety of emerging commercial and military applications such as aerial surveillance, border defense, traffic control, transportation, logistics, precision agriculture, search and rescue operations, disaster relief, to name a few, has sparked a lot of interest \cite{austin2011unmanned}.
Recently, many researches have studied the possibility of using (UAVs) as aerial base stations (ABSs) to support terrestrial users and assist terrestrial base stations (TBS) \cite{14,15,16,17,18,19}. This is motivated by the high probability of establishing a line-of-sight (LoS) channel when the BS is deployed at high altitude. Moreover, The UAV's mobility and relocation capability would greatly improve its deployment flexibility and give it the ability to intelligently adjust its positions in real-time in order to afford high quality of service (QoS). 

In addition, because of its simple, fast, and cost-efficient deployment, UAVs can be used in emergency and disaster recovery scenarios to serve mobile users in recovering areas~\cite{8644135}. It also can be used to provide extra coverage to areas that experience heavy traffic conditions exacerbated by mass events such as  large-scale activities such as sporting events, festivals, conferences, exhibitions and concerts.

Indeed, {UAV}s introduce several improvements over conventional infrastructure, but practical limitations have prevented it from attracting the attention of the industrial sector. The most important limitation is the power supply which extremely affect the flight time and the ability of communication and data processing. This reduces the efficiency of the UAVs since it needs to revisit a ground station every period of time to
recharge or change the battery and, as a consequence, its coverage area  will be temporarily out of service. Authors in \cite{7317490} present the issues of UAV communication networks in more details. 

Fortunately, these issues can potentially be solved using tethered UAV (T-UAV) which is a UAV connected to a ground station (GS) with a tether. The tether carries two cables, one for power and another for data. The GS can be located on a rooftop or a mobile station \cite{6G}.
With a stable power supply, the T-UAV can achieve much longer flight times, and support the required power for on-board communication and processing.  Beside the continuous energy supply, the tether allows   totally secured and high-speed data transfer (backhaul link).

Nevertheless, the T-UAVs are also susceptible to some limitations caused by the tether \cite{3-Dplacement}. First, the UAV is no more flexible. It depends on the GS's location, which depends on the availability and accessibility of rooftops.
Second, the tether of the T-UAV restrains the freedom of its mobility around the GS. In fact, besides the maximum tether length, the T-UAV is also constrained by the minimum inclination angle which is a security angle to avoid tangling with the buildings surrounding the GS and to ensure safety.

Despite these limitations, T-UAV  can be a viable alternative for untethered UAV (U-UAV) especially that it has been demonstrated in previous work that providing a sufficient number of GS locations accessibility and long enough tether, T-UAV outperforms untethered UAV \cite{optDep}. 

{\color{black} Unlike T-UAV, U-UAV has no mobility constraints. It can move around freely allowing us to deploy it in the optimal location, which is the cluster center as demonstrated in \cite{optDep}. 
However, U-UAV is limited by its energy. While T-UAV can stay in the air and operate without interruption for days, U-UAV can only hover for one or two hours before landing to change their batteries. Hence, U-UAV can not be continuously available like T-UAV. To account for U-UAV unavailability when analyzing U-UAV systems, researchers introduce a duty-cycle parameter $A \in [0 ; 1]$ that is determined by the charging and serving duration of the U-UAV \cite{optDep,6G}. Regarding the channel model, both T-UAV and U-UAV links experience Nakagami-m fading but with distinct parameters. Therefore, the primary distinctions between T-UAVs and U-UAVs come down to mobility restrictions and flight duration.

}

In this paper, we address a novel setup related to the large-scale deployment of tethered UAVs for wireless coverage enhancement. More details on the contributions of this paper are provided later in Sec. I-B. But first, we go over the most relevant literature in the following subsection.


\subsection{Related work}
Motivated by its tractability, stochastic geometry has been widely used to analyze integrated aerial-terrestrial networks \cite{3-Dplacement, optDep, Alzenad, Boris,8681266,zhao2020non,8888216,qin2021influence}. In \cite{Alzenad}, the authors suggested a method for analyzing coverage and rate in an aerial-terrestrial
network. They used PPP to model the locations of the BSs and the ABSs are assumed to be at the same altitude. The downlink performance is analyzed under the assumption that air to ground (A2G) and terrestrial links experience Nakagami-m fading and considering the occurrence of Line of sight (LoS) and Non Line of Sight (NLoS) transmissions separately. A slightly different scenario is considered in \cite{Boris} where the UAVs are assumed to be deployed above user hotspots. The influence of the correlation between the user hotspot locations and the UAV deployment is also studied.

Unlike most of existing literature, which focuses only on untethered {UAV}s, authors in \cite{3-Dplacement} and \cite{optDep} studied a communication system that uses tethered {UAV}s to provide wireless coverage. Both works focus mainly on the optimal deployment of the tethered {UAV} given the constraints imposed by the length of the tether and the safety measures taken to avoid tether entanglement with surrounding buildings. 
\cite{3-Dplacement} use a deterministic setup where it focuses on one user and one T-UAV to optimize the location of the T-UAV by minimizing the path loss, while \cite{optDep} focuses on one cluster where users are uniformly distributed and determines the optimal placement by maximizing the coverage performance.  
In the same context, a more advanced solution was proposed in \cite{9739716}. In fact, authors in \cite{9739716}  model the optimal deployment problem by a Markov decision process and use reinforcement learning, more specifically a multi-agent Q-learning algorithm to solve it.

Other than the optimal placement, authors in \cite{8644135} use T-UAV to provide a high-quality backhaul connection to the ABSs that construct a flying network. Additionally, they use it to manage and control the ABS network.
More recently, authors in \cite{lou2021green} have proposed using tethered {UAV}s in a system architecture that aims to reduce the electromagnetic field (EMF) exposure of the users while maintaining their data rates. Moreover, in \cite{9420290}, T-UAVs were used to compensate the deficiency of TBSs  in rural areas and improve the coverage performance.

The most related works to this paper are \cite{3-Dplacement} and \cite{optDep}. However, unlike those two works, we study a large-scale T-UAV deployment system capturing large numbers of clusters and spatial distribution of users within each cluster. We use tools from stochastic geometry to analyze this system, and determine the optimal placement by minimizing the average path loss. We investigate the coverage performance of the system taking into consideration the interference, which was neglected in both \cite{3-Dplacement} and \cite{optDep}.

\subsection{Contributions}
\bigskip
Different from existing literature, which mainly focused on the performance of untethered {UAV}s, this paper presents the first attempt to analyze the performance of large scale deployment of tethered {UAV}s in a {UAV}-assisted wireless communication system. More details about the contributions of this paper are provided next.
\bigskip

{\em Tethered {UAV}s Large-Scale Deployment.} We consider a scenario in which tethered {UAV}s are needed to be deployed at user hotspots to assist TBSs in providing coverage. However, the deployment is limited by (i) accessible buildings that allow deployment of the {UAV} ground stations on their rooftops, (ii) maximum achievable length of the tether, and (iii) the surrounding environment which limits the inclination angle of the tether. We provide a mathematical framework that captures all these constraints, optimizes the location of each given {UAV} based on the distance between the hotspot center and its nearest accessible rooftop, and enables computing the overall coverage probability of the considered system as a function of all the tethered {UAV} system parameters, such as the tether length, the density of the hotspots, and the density of accessible buildings.

{\em Novel Stochastic Geometry-based Approach.} While stochastic geometry based analysis of the coverage of ABSs has already been studied before, the usual assumption is deploying all ABSs at the same altitude. Given that in the considered setup each tethered {UAV} will have its own altitude depending on the distance between the rooftop and the hotspot center, the equal altitude assumption can not hold. Hence, we propose a novel approach that divides each hotspot (user cluster)  into a finite number of  (rings) with each ring corresponding to a given optimal altitude. This enables quantizing the possible altitudes of the {UAV}s, and hence, enables analyzing the interference and the coverage of the considering system without the similar altitude assumption.

{\em System Level Insights.} Using numerical results of the derived expressions and Monte-Carlo simulations, we reveal multiple useful insights. For instance, we show that the maximum tether length of the {UAV} has an optimal value in an urban environment while increasing its value in a suburban environment actually reduces the coverage probability. We also learn that, unlike what is typically assumed in literature, it is actually not always the best approach to deploy a {UAV} for each user hotspot. In fact, there exists an optimal fraction of the hotspots for which we deploy tethered {UAV}s, which maximizes the coverage probability. We also learn that this 
optimal fraction reduces as we increase the fraction of accessible buildings.

\begin{table}[h]
\setlength\tabcolsep{2pt}
\centering
\caption{Summary of important notation.}\label{tab}
\begin{tabular}{c|c} 
\hline
\multirow{2}{*}{Notation} & \multirow{2}{*}{Description} \\ &    \\

\hhline{==}
\multicolumn{1}{c|}
{$R_0, N$} & the raduis of the cluster, the number of rings.\\
\hline
\multicolumn{1}{c|}{$\Phi_C, \lambda_C$}& PPP of cluster centers and its density.\\
\hline
\multicolumn{1}{c|}{$\Phi_{\rm TBS}, \lambda_T$}& PPP of TBS locations and its density.\\
\hline
\multicolumn{1}{c|}{$\delta$} & fraction of clusters in which T-UAVs are deployed.\\
\hline
\multicolumn{1}{c|}{$p_i$}& the probability of having the GS in ring $i$.\\
\hline
\multicolumn{1}{c|}{$p_{\rm out}$} & probability of having no ABS in a specific cluster.\\
\hline
\multicolumn{1}{c|}{$\rho_{\rm T}$, $\rho_{\rm ABS}$ }
&TBS and ABS transmission power.\\
\hline
\multicolumn{1}{c|}{$p_{L_i}, p_{N_i}$}& \multicolumn{1}{c}{\begin{tabular}[c]{@{}c@{}}the probability that an ABS associated with
a GS located in ring $i$ \\is a LoS/NLos ABS. \end{tabular}}\\
\hline
\multicolumn{1}{c|}{$m_L, m_N, m_T$}& Nakagami-m fading parameters.\\
\hline
\multicolumn{1}{c|}{$P^r_{N_i}$, $P^r_{L_i}$, $P^r_{T}$}&  received power from NLoS ABS, LoS ABS, TBS.\\
\hline
\multicolumn{1}{c|}{$\alpha_{\rm T}$, $\alpha_{\rm N}$, $\alpha_{\rm L}$}& the path-loss exponents for terrestrial, LoS and NLoS transmissions.\\
\hline
\multicolumn{1}{c|}{$\mu_{\rm T}$, $\mu_{\rm N}$, $\mu_{\rm L}$}&  the mean additional losses for terrestrial, LoS and NLoS transmissions.\\
\hline
\multicolumn{1}{c|}{$D_T, D_{L_i}, D_{N_i}$}& \multicolumn{1}{c}{\begin{tabular}[c]{@{}c@{}}distance from the typical user to the closest TBS, LoS ABS, and NLoS ABS\\ given that the GS associated to the ABS is located in ring $i$.\end{tabular}}\\
\hline
\multicolumn{1}{c|}{$R_{u_i,r}$, $D_{u_i,r}$}&
\multicolumn{1}{c}{\begin{tabular}[c]{@{}c@{}}the horizontal and euclidean distance between the typical user and the cluster\\ ABS associated with a GS located in ring $i$.\end{tabular}}\\
\hline
\multicolumn{1}{c|}{$d_{\rm Q1Q2}$}& \multicolumn{1}{c}{\begin{tabular}[c]{@{}c@{}}the minimum distance to the closest interfering $Q_2$ given that the typical user\\is associated with $Q_1$, $\{Q_1,Q_2\} \subset \{\rm LoS \ ABS,NLoS \ ABS,TBS\}$.\end{tabular}}  \\
\hline
\multicolumn{1}{c|}{$h_{u_i}$} & the ABS altitude given that it is associated to a GS located at ring $i$.\\
\hline
\multicolumn{1}{c|}{ $\mathbb{A}_{\rm L_i}, \mathbb{A}_{\rm N_i}, \mathbb{A}_{\rm T}$ } & \multicolumn{1}{c}{\begin{tabular}[c]{@{}c@{}} the association probabilities of LoS ABS, NLoS ABS, TBS  given that the GS\\ associated with the ABS is located at ring $i$.\end{tabular}}\\
\hline
\multicolumn{1}{c|}{ $\mathbb{A}_{\rm CL_i}, \mathbb{A}_{\rm CN_i}$ } & \multicolumn{1}{c}{\begin{tabular}[c]{@{}c@{}} the association probabilities of LoS and NLoS cluster ABS \\given that it is associated to a GS located in ring $i$.\end{tabular}}\\
\hline
\multicolumn{1}{c|}{ $L_I$ } & The Laplace transform of the interference.\\
\hline
\multicolumn{1}{c|}{ $P_{\rm cov}$ } & 
Average coverage probability.\\
\hline
\multicolumn{1}{c|}{$P^{\{L_i,N_i,CL_i,CN_i,T\}}_{\rm cond}$ } & \multicolumn{1}{c}{\begin{tabular}[c]{@{}c@{}}the conditional coverage probability given that the serving BS is LoS ABS,\\ NLoS ABS, LoS cluster ABS, NLoS cluster ABS, TBS.\end{tabular}}\\

\hline
\end{tabular}
\end{table}

\section{System model}\label{SysMod}

\begin{figure}[ht]
		\centering
		\includegraphics[scale=0.4]{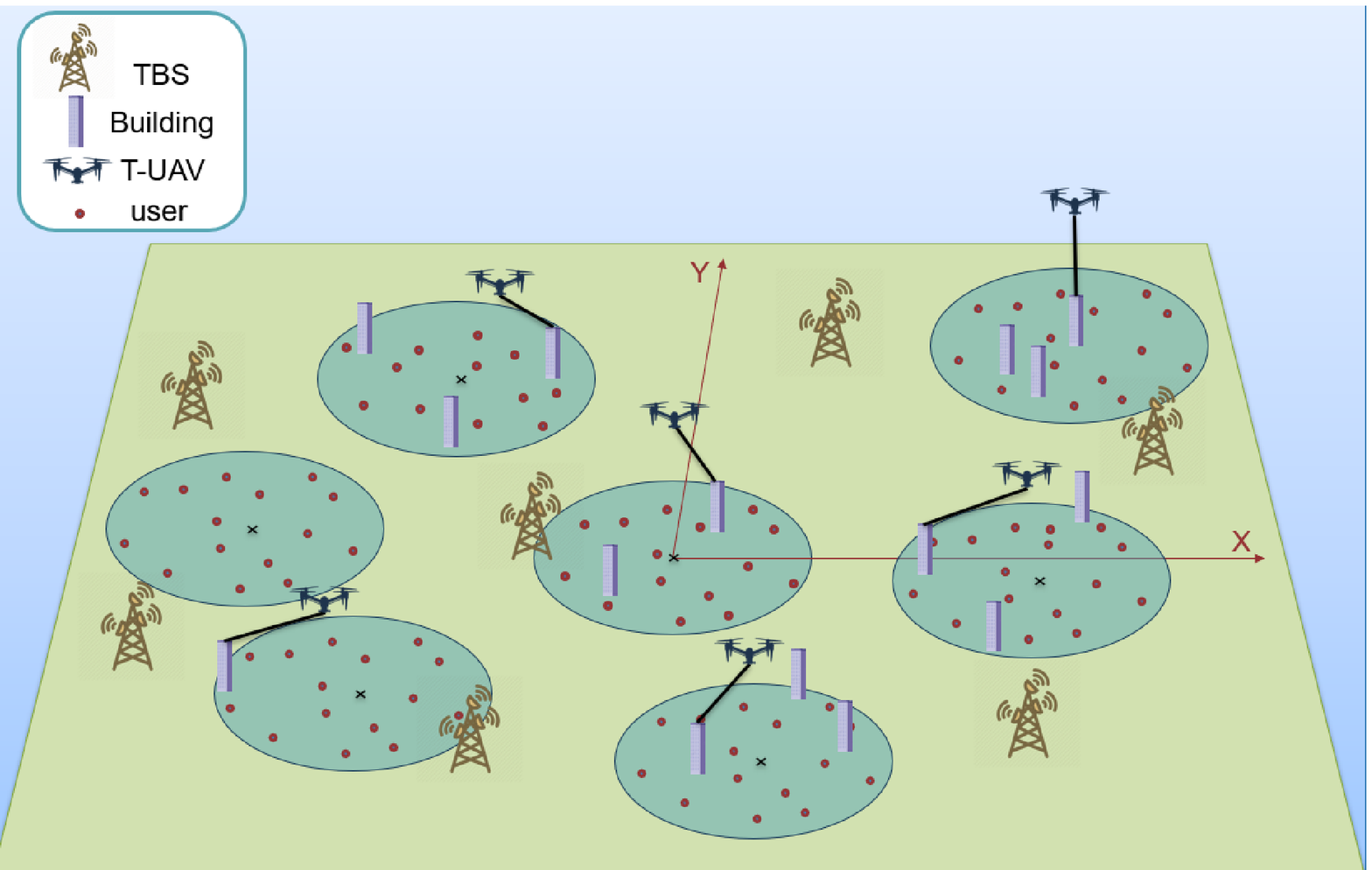}
		\caption{Illustration of the system model.}
		\label{gl}
\end{figure}
\subsection{Cluster model}

As depicted in Fig.\ref{gl}, we focus on a scenario in which users are located in high density hotspots (also referred to as clusters), where the locations of the users are uniformly distributed within each hotspot.
{\color{black} We will perform our analysis on a randomly selected user from a randomly selected cluster, which is referred to as the typical user.}
Depending on buildings availability, a T-UAV ground station is deployed on the nearest rooftop to the center of each hotspot. {For each cluster, at most one T-UAV is deployed.} {If no available building exists within the cluster, no T-UAV will be deployed in that cluster.}  To ensure generality of the proposed analysis, we assume that T-UAVs are deployed at only a fraction $\delta$ of the total number of clusters. A hotspot region is modeled as a disk $D(L_0, R_0)$ centered at the origin $L_0$ with radius $R_0$ and the set of hotspot centers is a homogeneous Poisson Point Process (PPP), denoted by $\Phi_{C}$ with intensity $\lambda_C$. Since we only deploy ABS in a specific percentage of clusters,  the density of T-UAV equipped cluster centers is $\delta\lambda_C$.

{
Given that we can have a rooftop at any location in the cluster, we can have infinite possible values of the distance between the centre of the cluster and the rooftop. Thus, to make our analysis more tractable, we quantify the locations of rooftops by dividing each cluster into $N$ frames (concentric rings).
For each rooftop location we have an optimal placement for the T-UAV and hence for each location we have an optimal altitude. Thus discretizing the distance between the rooftop and the centre of the cluster allows us to avoid deploying T-UAVs at a particular height (generally the average of UAVs' height) as has been done in previous works such as \cite{Alzenad}.} {So, unlike untethered UAVs that are flexible and free to move, 
the placement of T-UAVs depends directly on the location of the ground station.}  


As a result and using thinning procedure {(Theorem 3.3 in \cite{mukherjee2014analytical})},  the Poisson Point Process of the cluster centers $\Phi_{C}$ can be decomposed into $N$ PPP $\left \{ \Phi_{C_i} \right \}_{i \in \left \{ 1,..,N \right \}}$ where $\Phi_{C}=\bigcup_{i \in \left \{ 1,..,N \right \}}\left \{ \Phi_{C_i} \right \}$. If the nearest rooftop is located in the ring $i$, the cluster center is mapped into $\Phi_{C_i}$. So, the intensity of $\Phi_{C_i}$ is $\lambda_{C_i}=p_i\delta\lambda_{C}$ where $p_i$ is the probability of having the nearest rooftop in the ring $i$.

The cluster centers and the users can be modeled as a Matern cluster process where the parent is the cluster center and the daughter points are the users which are uniformly scattered on the disk of radius $R_0$ centred at each parent point. The intensity of this Matern cluster process is $\lambda=\delta\lambda_C\Bar{c}$ where $\Bar{c}$ is the mean number of daughter points per parent. Using thinning procedure, we can get $N$ Matern cluster processes.

{


}

\subsection{Network model}
\subsubsection{TBS}
The locations of the TBSs are modeled as a homogeneous PPP $\Phi_{TBS}$ with intensity $\lambda_{\rm T}$ and all the TBSs transmit at the same power $\rho_{\rm T}$. {The height of TBSs is assumed to be equal to zero. The same applies for the users. This is commonly used in literature to focus on the influence of the ABS heights \cite{Alzenad}.}

\subsubsection{ABS}
All the ABSs transmit at the same power $\rho_{\rm ABS}$ and using the displacement theorem, we can model their locations as the displacement of the locations of the cluster centers, due to the randomness of the distances between the rooftops and their corresponding cluster centers. 
We denote the PPP of the locations of ABSs associated with GSs located in ring $i$  by $\Phi_{\rm ABS_i}$.  
Mathematically, we define $\Phi_{\rm ABS_i}$ as 
$
    \Phi_{\rm ABS_i}=\left\{x\in \Phi_{C_i}: x+V_x \right \},
$
where the random variables $V_x$ are independent. $V_x$ is the displacement caused by the T-UAV being connected to a rooftop that is not located at the cluster center. Since $\Phi_{C_i}$ is a PPP, $\Phi_{\rm ABS_i}$ is also a PPP with intensity
\begin{equation}
    \lambda_{\rm ABS_i}=p_i\delta\lambda_C\int_{\mathbb{R}^2}\vartheta (x,y){\rm d}x.
\end{equation}

$\vartheta (x, y)= \frac{\mathbbm{1}_{C(x,R_{u_i})}(y)}{2\pi R_{u_i}}=\frac{\mathbbm{1}_{C(o,R_{u_i})}(y-x)}{2\pi R_{u_i}}$ is the PDF of the new location y {where $\mathbbm{1}_{C(x,R_{u_i})}(y)=1$ if $y \in C(x,R_{u_i}) $ and $\mathbbm{1}_{C(x,R_{u_i})}(y)=0$ otherwise}. It displaces each point by a random vector uniformly on $C(o,R_{u_i})$ where $C(o,R_{u_i})$ is the circle of center $o$ and radius $R_{u_i}$, and $R_{u_i}$ is the horizontal distance between an ABS associated with GS located in ring $i$ and the center of the cluster where the ABS is deployed.

Since $\left\{\Phi_{C_i}\right\}_{i \in \left[1,N\right]}$ are homogeneous PPPs and $\vartheta(x,y)$ is a function only of $y-x$, we have
\begin{equation}
    \lambda_{\rm ABS_i}=p_i\delta\lambda_C , \ \forall i \in \left[1,N\right].
\end{equation}

In addition, air to ground (A2G) links can be either a Line of sight (LoS) links or Non Line of Sight (NLoS) links. This depends on the nature of the surrounding environment. The probability of LoS transmission, denoted by $p_L$ and provided in \cite{Los}, is defined as 
\begin{equation}
    p_L(r,h)=\frac{1}{1+a \exp({-b(\frac{180}{\pi} \tan^{-1}(\frac{h}{r})-a})},
\end{equation}
where $h$ is the T-UAV altitude, $r$ is the horizontal distance between the typical user and the projection of the T-UAV on the horizontal plane, $a$ and $b$ are constants that depend on the environment. Consequently, the probability of NLoS transmission, denoted by $p_N$ is defined as $p_N(r, h) = 1-p_L(r, h)$.

So, using thinning procedure, the PPP $\Phi_{\rm ABS_i}$ can be decomposed into two PPPs $\Phi_{\rm L_i}$ and $\Phi_{\rm N_i}$ with densities $\lambda_{\rm L_i}$ and $\lambda_{\rm N_i}$, respectively. 

An ABS's location is mapped into $\Phi_{\rm L_i}$, if the ABS is in a LoS condition with the typical user and the GS associated with this ABS is located in the ring $i$ while it is mapped into $\Phi_{\rm N_i}$ if it is in a NLoS condition with the typical user and the GS associated to it is located in the ring $i$.

Since it can be changed from one ring to another, we denote the altitudes of ABSs associated with GSs located in ring $i$ by $h_{u_i}$. We also denote the probability that an ABS associated with a GS located in ring $i$ is a LoS ABS by $p_{L_i}$ which is defined as $p_{L_i}(r)=p_L(r,h_{u_i})$. Similarly, we denote $p_{N_i}(r)=1-p_{L_i}(r)$.

To sum up, ABS is modeled as a PPP $\Phi_{\rm ABS}$ which is decomposed into $2N$ PPP $\left \{ \Phi_{\rm L_i} \right \}_{i \in\left[1,N\right]}$ and $\left \{ \Phi_{\rm N_i} \right \}_{i\in\left[1,N\right]}$ with intensities 
\begin{equation}
    \left\{\begin{matrix}
    \lambda_{\rm L_i}(r)=p_{L_i}(r)\lambda_{\rm ABS_i}=p_{L_i}(r)p_i\delta\lambda_C , && \forall i\in\left[1,N\right],\\ 
    \lambda_{\rm N_i}(r)=p_{N_i}(r)\lambda_{\rm ABS_i}=p_{N_i}(r)p_i\delta\lambda_C , && \forall i\in\left[1,N\right].
    \end{matrix}\right .
\end{equation}

\subsubsection{GS}\label{GS}
The locations of the rooftops that are capable of hosting (also referred to as accessible rooftops) the T-UAV ground stations (GSs) are modeled as a homogeneous PPP $\Phi_{\rm GS}$ with intensity $\lambda_{\rm GS}=\kappa_b \lambda_b$ where $\kappa_b $ is the accessibility factor, which is the fraction of buildings that have a rooftop eligible for GS deployment,  and $\lambda_b$ is the density of the buildings.

For a given cluster center,  the probability of having the nearest rooftop in the ring $i$ is defined 

\begin{equation}
\text{as} \quad
    p_i= {\mathbb{P}(R_{f_{i-1}}<R_{\rm GS}<R_{f_{i}} )} 
    ={\exp(-\lambda_{\rm GS}\pi R_{f_{i-1}}^2)\left(1- \exp\left(-\lambda_{\rm GS}\pi (R_{f_i}^2-R_{f_{i-1}}^2)\right) \right)},
\end{equation}

where $R_{f_i}=\frac{R_0}{N}i$, $i \in\left[0,N\right]$ and $R_{\rm GS}$ is the distance between the cluster center and the projection of the GS.

In addition, the probability of having no accessible rooftops inside the cluster is given by
\begin{equation}
    p_{out}= \mathbb{P}(R_0<R_{GS} )
    =\exp(-\lambda_{\rm GS}\pi R_0^2).
\end{equation}

\subsection{Channel model}
We assume that A2G and terrestrial links experience Nakagami-m fading with shape parameters and scale parameters, given by $(m_{\rm N},\frac{1}{m_{\rm N}})$, $(m_{\rm L},\frac{1}{m_{\rm L}})$ and $({m_{\rm T}},\frac{1}{m_{\rm T}})$ for NLoS, LoS and terrestrial links, respectively. Note that Rayleigh fading can be generated from Nakagami-m fading by setting the shape parameter to unity.
The probability density function of the channel fading power gains, denoted by $G$, is defined as
\begin{equation}
    f_{G_Q}(g)=\frac{m_Q^{m_Q}.g^{m_Q-1}}{\Gamma(m_Q)}e^{-m_Q.g} ,  Q\in \left \{\rm N, L, T \right \},
\end{equation}
where $\Gamma(m_Q)$ is the Gamma function given by $\Gamma(m_Q)=\int_{0}^{\infty }x^{m_Q-1} e^{-x} dx$.

The received power at the typical user from a LoS ABS associated with GS located in ring $i$, NLoS ABS associated with GS located in ring $i$ or, TBS is given by

\begin{equation}
    P^r_{\rm N_i}=\rho_{\rm ABS}\eta_{\rm N}G_{\rm N}D_{\rm N_i}^{-\alpha_{\rm N}},\quad P^r_{\rm L_i}=\rho_{\rm ABS}\eta_{\rm L}G_{\rm L}D_{\rm L_i}^{-\alpha_{\rm L}},\quad P^r_{\rm T}=\rho_{\rm T}\eta_{\rm T}G_{\rm T}D_{\rm T}^{-\alpha_{\rm T}},
\end{equation}
where $D_{\rm N_i}$, $D_{\rm L_i}$ and $D_{\rm T}$ are the distances from  the typical user to NLoS ABS associated with GS located in ring $i$, LoS ABS associated with GS located in ring $i$ and TBS, respectively, 
$\alpha_{\rm T}$, $\alpha_{\rm L}$ and $\alpha_{\rm N}$ are the path-loss exponents for terrestrial, LoS and NLoS transmissions ($\alpha_{\rm L} \leq \alpha_{\rm N}$), and $\eta_{\rm T}$,  $\eta_{\rm L}$ and $\eta_{\rm N}$ are the mean additional losses for terrestrial, LoS and NLoS transmissions ($\eta_{\rm L} \geq \eta_{\rm N}$). 
 
To measure the quality of the downlink between the typical user and its serving BS, the signal-to-interference plus noise ratio (SINR) is used. The instantaneous SINR at the typical user when it is associated with a BS located at $z_0$ is defined as

\begin{equation}
    {\rm SINR}=
    \left\{\begin{matrix}
    \frac{P^r_{\rm T,z_0}}{\sigma^2+I} &if z_0 \in \Phi _{\rm T}\\ 
    \frac{P^r_{\rm L_i,z_0}}{\sigma^2+I} &if z_0 \in \Phi_{\rm L_i}, \forall i\in\left [ 1,N \right ]\\
    
    \frac{P^r_{\rm N_i,z_0}}{\sigma^2+I} &if z_0 \in \Phi_{\rm N_i}, \forall i\in\left [ 1,N \right ]
    \end{matrix},\right.
\end{equation}

where $\sigma^2$ is the additive white Gaussian noise power and $I$ is the interference power. The ABSs and TBSs are assumed to operate in the same frequency spectrum  and thus, they interfere with each other. We also assume the worst-case scenario where all BSs has data to send. In this case, the interference is equal to the sum of  the received powers from all the BSs except the serving BS. It is formulated mathematically as
\begin{equation}\label{interference}
    I=\sum_{y\in\Phi_{\rm T}\setminus\left\{z_0\right\}}P^r_{\rm T,y} + \sum_{i\in\left[1,N\right]}\sum_{y\in\Phi_{\rm L_i}\setminus\left\{z_0\right\}}P^r_{\rm L_i,y} + \sum_{i\in\left[1,N\right]}\sum_{y\in\Phi_{\rm N_i}\setminus\left\{z_0\right\}}P^r_{\rm N_i,y} + P^r_{\rm CABS},
\end{equation}
where $P^r_{\rm CABS}$ is the received power from the ABS located in the same cluster as the typical user. It is equal to zero, if there is no ABS in the same cluster as the typical user.

\subsection{Association Policy}

To determine the serving BS, we adopt the strongest average received power association scheme. Hence, having a NLoS ABS closest to the typical user does not mean that it delivers the strongest average received power. A LoS ABS may provide a stronger average received power due to the fact that $\eta_{\rm L} > \eta_{\rm N}$ and $\alpha_{\rm L} < \alpha_{\rm N}$. But, for a particular set of BS (LoS ABS, NLoS ABS, or TBS), the transmit powers and path-loss parameters are the same. So, the closest BS provides a higher average received power than that provided by any other BS in its set. Finally, we conclude that the serving BS can be either the closest NLoS ABS, the closest LoS ABS, or the closest TBS.

\section{Optimal deployment of TUAV}

{\color{black}
Users are uniformly distributed inside clusters. They are continuously in motion so on average the optimal location of the UAV is above the centre of the cluster, as has been demonstrated and proven in \cite{optDep}. Since T-UAVs are constrained by the position of the closest accessible building where the GS will be placed, the tether length, and the inclination angle, we are unlikely to be able to place them above the cluster's center. T-UAVs will be deployed in the most advantageous position that they can reach.
}
Hence, the ultimate goal of T-UAV placement is to find the location that minimizes the average path loss between the T-UAV and the users within its cluster.

\begin{figure}[ht]
		\centering
		\includegraphics[scale=0.6]{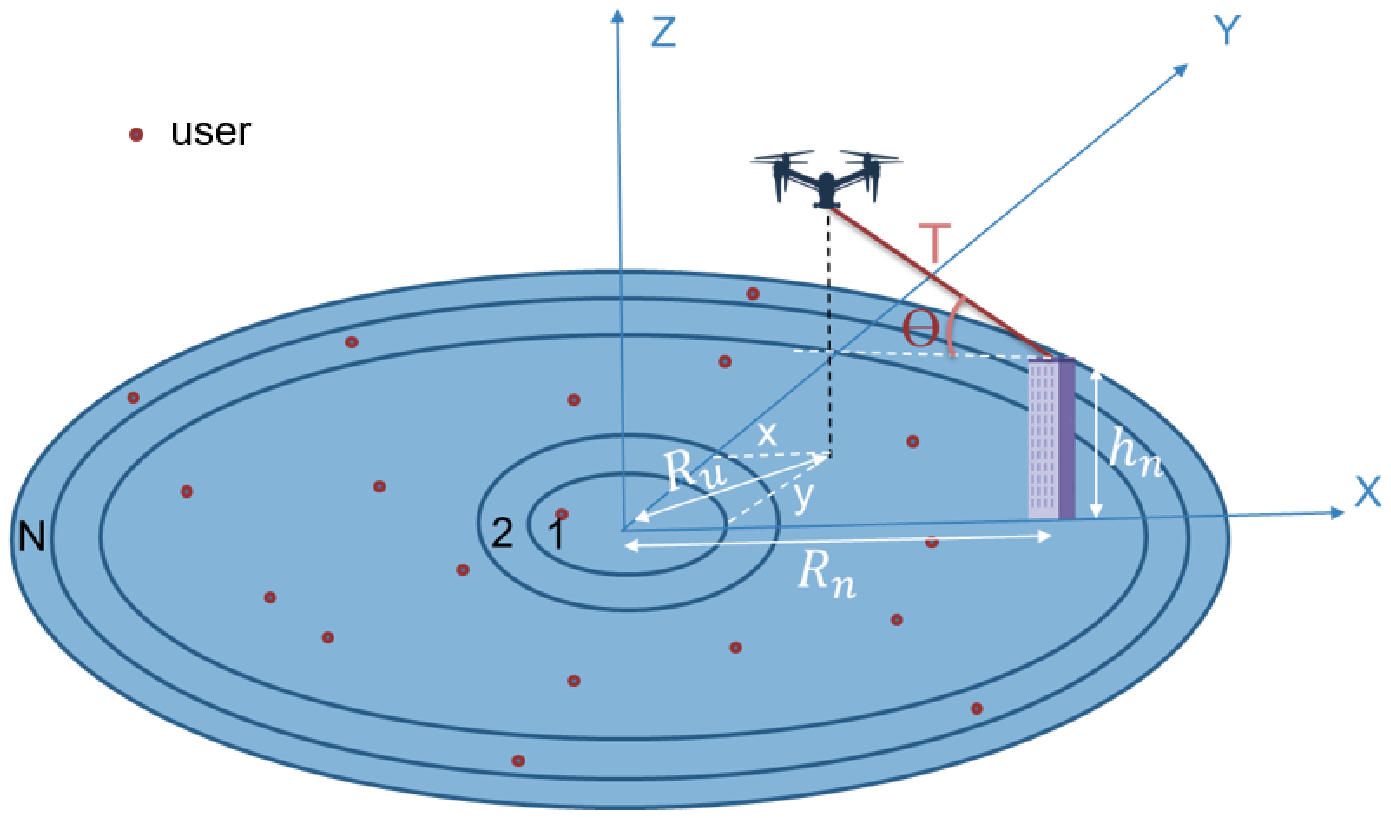}
		\caption{The system setup considered in this section.}
		\label{1}
\end{figure}

We will perform our analysis for a randomly selected hotspot region.
As shown in Fig.\ref{1}, we consider a system composed of a T-UAV launched from a GS that is placed on a rooftop at height $h$. The T-UAV has the freedom to hover anywhere within the hemisphere centered at the rooftop with radius equal to the maximum value of the tether length $T_{\rm max}$. In order to avoid tangling upon surrounding buildings and to ensure safety of the tether, the inclination angle of the tether, has a minimum value $\theta_{\rm min}$.
To represent the locations of the T-UAV and the GS, We use Cartesian coordinates. So we denote the location of the T-UAV as $(x, y, h)$ and the location of the GS as $(R_n,0,h_n)$ where $n$ is the index of the ring in which the GS is located. The  horizontal distance between the GS and the center of the cluster is defined as $R_n=\frac{2n-1}{2N}R_0$ since we approximate the placement of the buildings to the center of the ring in which they are located. We assume that all the buildings have the same altitude $h_n$, regardless the ring in which they are situated.
The optimization problem can be formally written as follows.
\begin{equation}\label{optpb}
    \begin{split}
         \min_{T,\theta } {\rm PL}_{\rm avg}(&T,\theta) \\
         \text{subject to: }&\theta_{\rm min} \leq \theta \leq \frac{\pi}{2}\\
         &0 \leq T \leq T_{\rm max},
    \end{split}
\end{equation}
where ${\rm PL}_{\rm avg}$ is the average path loss that will be defined later in this section, {{$T$ is the tether length and $\theta$ is the inclination angle.}} 
So, having $T_{\rm max}, \theta_{\rm min}, h_n$ and $R_n$, we can determine the optimal values of the tether length  and the tether inclination angle  by minimizing the average path loss. Since $R_n$ is proportional to the ring index ($n$), the optimization problem will be solved N times for a given $T_{\rm max}, \theta_{min}$ and $ h_n$.

The path loss between the T-UAV and a typical user can be calculated using the distance $d$ between the typical user and the T-UAV as well as the transmission parameters. It is  defined as 

\begin{equation}
    {\rm PL}=\frac{1}{\eta_{\rm N}}p_N(\sqrt{d^2-h^2},h) d^{\alpha_{\rm N}}+ \frac{1}{\eta_{\rm L}}p_L(\sqrt{d^2-h^2},h) d^{\alpha_{\rm L}}.
\end{equation} 

To solve the optimization problem (\ref{optpb}), we need to write the expression of the path loss as a function of the tether length $T$ and the tether inclination angle $\theta$.

The T-UAV altitude can be written in function of $T, \theta$ and $h_n$ as
\begin{equation}\label{4.3}
    h=h_n+T\sin (\theta).
\end{equation}

So, the Euclidean distance between the typical user and the T-UAV can be written as 
\begin{equation}\label{4.4}
    d=\sqrt{R_{u,r}^2+h^2}=\sqrt{R_{u,r}^2+(h_n+T\sin (\theta))^2},
\end{equation}
where $R_{u,r}$ is the  horizontal distance between the typical user and the T-UAV.

The probability density function (PDF) of $R_{u,r}$, provided in \cite{optDep}, is 
\begin{equation}
    f_{R_{u,r}}(r)= 
\left\{\begin{matrix}
\frac{2r}{R_0^2} & 0\leq r\leq R_0-R_u\\ 
\frac{2r}{\pi R_0^2} \arccos(\frac{R_u^2+r^2-R_0^2}{2R_ur}) \qquad & R_0-R_u \leq r \leq R_0+R_u
\end{matrix},\right.
\end{equation}
where $R_u$ is the  horizontal distance between the center of the cluster and the T-UAV. It is defined 
\begin{equation}
\text{as} \qquad \qquad \qquad \qquad \qquad 
    R_u= \sqrt{R_n^2-2R_n \sqrt{T^2\cos^2(\theta)-y^2}+T^2\cos^2(\theta)}. \qquad  \qquad 
\end{equation}

{As discussed before, the optimal placement of a T-UAV is as close to the cluster center as possible. So, considering the line connecting the cluster centre and the building position as the x axis, the optimal value of y is zero. As a result, the distance $R_u$ can be reduce to  } 
\begin{equation}
    R_u=R_n-T \cos(\theta).
\end{equation}

Using (\ref{4.4}), we can write the expression of the path loss as a function of $T, \theta$ and the horizontal distance between the typical user and the T-UAV denoted by $R_{u,r}$.
This expression is given by

\begin{equation}
    \begin{split}
        {\rm PL}(T,\theta,R_{u,r})=&\frac{p_N(R_{u,r},h)}{ \eta_{\rm N}}   \sqrt{R_{u,r}^2+(h_n+T\sin (\theta))^2}^{\alpha_N}\\
        &+
        \frac{p_L(R_{u,r},h)}{ \eta_{\rm L}} \sqrt{R_{u,r}^2+(h_n+T\sin (\theta))^2}^{\alpha_L}.
    \end{split}
\end{equation}

Now, we can compute the average path loss as follows 
\begin{equation}
    {\rm PL}_{\rm avg}(T,\theta)=  \mathbb{E}_{R_{u,r}} [{\rm PL}]=\int_0^{R_0+R_u} {\rm PL}(T,\theta,r) f_{R_{u,r}}(r)\ {\rm d}r,
\end{equation}
and solve the optimization problem (\ref{optpb}) to get the optimal value of the tether length $T_{\rm opt}$ and the optimal value of the tether inclination angle $\theta_{\rm opt}$. As we already mentioned, we have $N$ optimal values for each of the tether length and the tether inclination angle. Using (\ref{4.3}), we can deduce $N$ optimal values for the altitude of T-UAV. {So, unlike normal UAVs which are totally free to localize themselves in an optimal position the T-UAV will have $N$ different optimal 3D placement depending on the position of the GS.}
{\color{black} 
Its important to mention that we will not provide the mathematical solutions of the optimization problem (\ref{optpb}) since its hard to solve analytically. We will just solve it numerically using Matlab.}

\section{Coverage probability}

As mentioned earlier, the considered system is composed of ABSs deployed at a subset of user clusters (hotspots), TBSs, and users uniformly distributed inside the clusters. The typical user is randomly selected from a random cluster. Assuming there exists at least one accessible building in the cluster to deploy a GS, there exists an ABS in this cluster responsible for serving its users. Supposing that it is located in the point $(x_0,y_0)\in \Phi_{\rm ABS}$.
We partition the set $\Phi_{\rm ABS}$ into the sets $\left \{ (x_0,y_0) \right \}$ and $\Phi_{\rm ABS}\setminus\left \{ (x_0,y_0) \right \}$, containing the ABS located in the same cluster as the typical user and all the remaining ABSs, respectively.
Note that, following Slivnyak's theorem, the set of UAVs remains a PPP with intensity $\lambda_{\rm ABS}$. So, we can keep the same notation as those existing in Section \ref{SysMod}.
\subsection{Distance distributions}

The cumulative distribution functions (CDFs) and the probability density functions (PDFs) of the distances between the typical user and  the closest BS from each type (TBS, LoS ABS, and NLoS ABS) are crucial in the covarage analysis. They are needed to derive each of the association probabilities, the Laplace transform of interference, and even the final expression of the coverage probability.

\begin{lemma}\label{Lm1}
We denote the distances between the typical user and the closest LoS ABS associated with a GS located in ring $i$, the closest NLoS ABS associated with a GS located in ring $i$ and 

the closest TBS by $D_{\rm L_i}$, $D_{\rm N_i}$ and $D_{\rm T}$, respectively.  the CDF of these distances are given by

\begin{equation}\label{23}
    \left\{\begin{matrix}
    F_{D_{\rm T}}(d)=1-\exp\left(-\pi \lambda_{\rm T}d^2 \right)\mathbbm{1}_{[0,+\infty[}(d).\\ 
    F_{D_{\rm L_i}}(d)=\left[1-\exp\left(-2\pi p_i\delta\lambda_C \int_{0}^{\sqrt{d^2-h_{u_i}^2}}rp_{L_i}(r){\rm d} r\right)\right]\mathbbm{1}_{[h_{u_i},+\infty[}(d).\\ 
    F_{D_{\rm N_i}}(d)=\left[1-\exp\left(-2\pi p_i\delta\lambda_C \int_{0}^{\sqrt{d^2-h_{u_i}^2}}rp_{N_i}(r){\rm d}r\right)\right]\mathbbm{1}_{[h_{u_i},+\infty[}(d).
    \end{matrix}\right.
\end{equation}

\begin{IEEEproof}
The above results follow by using the null probability of a PPP given that the locations of TBSs are homogeneous PPP with density ${\lambda_T}$ while those of LoS (NLoS) ABSs follow an inhomogeneous PPP with density $p_i\delta\lambda_C p_{L_i}(r)$.
\end{IEEEproof}
\end{lemma}

\begin{cor}
    For each $i \in \left [ 1,N \right ]$, the probability density functions (PDFs) are
    \begin{equation}
        \left\{\begin{matrix}
        f_{D_{\rm T}}(d)=2 \pi \lambda_{\rm T}d \exp\left(-\pi \lambda_{\rm T}d^2 \right)\mathbbm{1}_{[0,+\infty[}(d).\\ 
        f_{D_{\rm L_i}}(d)=2\pi p_i \delta\lambda_C d p_{L_i}(\sqrt{d^2-h_{u_i}^2}) \exp\left(-2\pi p_i\delta\lambda_C \int_{0}^{\sqrt{d^2-h_{u_i}^2}}rp_{L_i}(r){\rm d}r\right)\mathbbm{1}_{[h_{u_i},+\infty[}(d).\\ 
        f_{D_{\rm N_i}}(d)=2\pi p_i \delta\lambda_C d p_{N_i}(\sqrt{d^2-h_{u_i}^2}) \exp\left(-2\pi p_i\delta\lambda_C \int_{0}^{\sqrt{d^2-h_{u_i}^2}}rp_{N_i}(r){\rm d}r\right)\mathbbm{1}_{[h_{u_i},+\infty[}(d).
        \end{matrix}\right.
    \end{equation}
    
    \begin{IEEEproof}
    The results follow directly by taking the derivative
    of the expressions in (\ref{23}).
    \end{IEEEproof} 
\end{cor}

After determining the distance distributions of TBS, LoS ABS and NLoS ABS, it is time to determine the distance distribution of the ABS located in the same cluster as the typical user. In this paper, we will refer to this ABS as the cluster ABS. 

\begin{theorem}\label{thm1}
The CDF of the horizontal distance between the typical user and the cluster ABS, given that the cluster ABS is associated with a GS located in ring $j$, is defined as

\begin{equation}\label{5.13}
    F_{R_{u_j,r}}(r) =
    \left\{\begin{matrix}
    \frac{r^2}{R_0^2},& \quad 0\leq r \leq R_0-R_{u_j}\\ 
    A,&  \quad R_0-R_{u_j}\leq r \leq R_0+R_{u_j} 
    \end{matrix},\right.
\end{equation}
where $A=\frac{\theta_2}{\pi}+\frac{1}{2\pi R_0^2}\left( -2r^2\arcsin(\frac{\sin(\theta_2)R_{u_j}}{r}) - 2\sin(\theta_2)R_{u_j}\sqrt{g(r,\theta_2)}  \right) +\frac{1}{2\pi R_0^2} (- \sin(2\theta_2)R_{u_j}^2 - 2\theta_2r^2 + 2\pi r^2)$ ,$\theta_1=\arcsin(\frac{r}{R_{u_j}})$, $\theta_2=\arccos(\frac{R_0^2+R_{u_j}^2-r^2}{2R_0R_{u_j}})$, and $g(r,\theta)=r^2-R_{u_j}^2\sin^2(\theta)$. $R_{u_j}$ is the horizontal distance between the center of the cluster and the ABS associated with GS located in ring $j$. It is defined as $R_{u_j}=R_j-T_j\cos(\theta_j)$ where $R_j=\frac{2n-1}{2N}R_0$ is the horizontal distance between the GS and the center of the cluster, $T_j$ is the optimal tether length when the GS is located in ring $j$, and $\theta_j$ is the optimal inclination angle when the GS is located in ring $j$.
\begin{IEEEproof}
See Appendix~\ref{app:thm1}.
\end{IEEEproof}
\end{theorem}

\begin{cor}
The PDF of the horizontal distance between the typical user and the cluster ABS, given that the cluster ABS is associated with a GS located in ring $j$, is given by

\begin{equation}
    f_{R_{u_j,r}}(r)=\frac{\partial F_{R_{u_j,r}}(r)}{\partial r}=\left\{\begin{matrix}
    \frac{2r}{R_0^2}, & 0\leq r < R_0-R_{u_j}\\ 
    D, & R_0-R_{u_j}\leq r \leq R_0+R_{u_j}\\
    0, & otherwise
    \end{matrix},\right.
\end{equation}
 
\begin{equation*}
    \begin{split}
    \text{where} \quad
        D =& \frac{1}{2\pi R_0^2}\left(\frac{-2r^2\left(  \frac{R_0^2+R_{u_j}^2-r^2}{2R_0^2R_{u_j}  \sin(\theta_2)} - \frac{R_{u_j} \sin(\theta_2) }{r^2}    \right)}{   \sqrt{1-\frac{R_{u_j}^2  \sin^2(\theta_2) }{r^2} }  }
        -4r \arcsin\left(\frac{R_{u_j}}{r} \sin(\theta_2) \right) \right)\\
        &+\frac{1}{2\pi R_0^2}\left( \frac{-r(R_0^2+R_{u_j}^2-r^2) \sqrt{r^2-R_{u_j}^2\sin^2(\theta_2)} }{R_0^2R_{u_j} \sin(\theta_2) }
         - \frac{\left(2r -  \frac{r(R_0^2+R_{u_j}^2-r^2)}{R_0^2 }  \right)\sin(\theta_2)}{ \sqrt{\frac{r^2}{R_{u_j}^2}-\sin^2(\theta_2)} } \right)\\
         &+ \frac{1}{2\pi R_0^2}\left( \frac{-2R_{u_j}r\cos(2 \theta_2) }{R_0 \sin(\theta_2)}- 4r\theta_2 - \frac{2r^3}{R_0 R_{u_j} \sin(\theta_2)}+4\pi r \right)+ \frac{r}{\pi R_0 R_{u_j}\sin(\theta_2)}.
    \end{split}
\end{equation*}
\begin{IEEEproof}
    The results follow directly by taking the derivative
    of the expressions in (\ref{5.13}).
    \end{IEEEproof} 
\end{cor}

Based on the distribution of the horizontal distance between the typical user and the cluster ABS, we can derive the distribution of the euclidean distance between the typical user and the cluster ABS. 

\begin{cor}
The CDF of the euclidean distance between the typical user and the cluster ABS is defined as
\begin{equation}
     F_{D_{u_j,r}}(d)=F_{R_{u_j,r}}\left(\sqrt{d^2-h_{u_j}^2}\right)\mathbbm{1}_{[h_{u_j},+\infty[}(d).
\end{equation}
\begin{IEEEproof}It follows  from the relationship between the euclidean and horizontal distances.
\end{IEEEproof}
\end{cor}

\begin{cor}
The PDF of the euclidean distance between the typical user and the cluster ABS is given by

\begin{equation}\label{PDF}
     f_{D_{u_j,r}}(d)=\frac{\partial F_{D_{u_j,r}}(d)}{\partial d}= \frac{d}{\sqrt{d^2-h_{u_j}^2}}f_{R_{u_j,r}}\left(\sqrt{d^2-h_{u_j}^2}\right)\mathbbm{1}_{[h_{u_j},+\infty[}(d).
\end{equation}
\end{cor}

\subsection{Nearest interfering base stations}
Before deriving the association probability expressions, we need first to determine the distances to the nearest interfering base stations when a TBS is the serving base station, when  a LoS ABS is the serving base station, and when  a NLoS ABS is the serving base station.
So, we provide the following distances. 

\begin{nrem}\label{RQ}
We denote $d_{Q_1Q_2}(r)$ the minimum distance to the closest interfering $Q_2$ given that the typical user is associated with $Q_1$ where $r$ is the distance between the serving BS and the typical user and $\left \{ Q_1,Q_2 \right \}\subset \left \{\rm L,N,T \right \}$. The index $i$ existing in some distances refer to the ring index in which a GS associated with an ABS is located. 
Those distances are given by
\begin{align}
    d_{\rm TL_i}(r)= 
    \left\{\begin{matrix}
    h_{u_i}, &\quad \text{if } r<l^i_{\rm T-L}\\ 
    \left( \frac{\rho_{\rm ABS}\eta_{\rm L}}{\rho_{\rm T}\eta_{\rm T}} \right)^{\frac{1}{\alpha_{\rm L}}}r^{\frac{\alpha_{\rm T}}{\alpha_{\rm L}}}, &\quad \text{if } r\geq l^i_{\rm T-L}
    \end{matrix}\quad,\right.\quad\quad&  d_{\rm LT}(r)= \left( \frac{\rho_{\rm T}\eta_{\rm T}}{\rho_{\rm ABS}\eta_{\rm L}} \right)^{\frac{1}{\alpha_{\rm T}}}r^{\frac{\alpha_{\rm L}}{\alpha_{\rm T}}},\\
    d_{\rm TN_i}(r)= 
    \left\{\begin{matrix}
    h_{u_i}, &\quad \text{if }  r<l^i_{\rm T-N}\\ 
    \left( \frac{\rho_{\rm ABS}\eta_{\rm N}}{\rho_{\rm T}\eta_{\rm T}} \right)^{\frac{1}{\alpha_{\rm N}}}r^{\frac{\alpha_{\rm T}}{\alpha_{\rm N}}}, &\quad \text{if } r\geq l^i_{\rm T-N}
    \end{matrix}\quad,\right. \quad\quad & d_{\rm NL}(r)= \left( \frac{\eta_{\rm L}}{\eta_{\rm N}} \right)^{\frac{1}{\alpha_{\rm L}}}r^{\frac{\alpha_{\rm N}}{\alpha_{\rm L}}},\\
    d_{\rm LN_i}(r)= 
    \left\{\begin{matrix}
    h_{u_i}, &\quad \text{if }  r<l^i_{\rm L-N}\\ 
    \left( \frac{\rho_{\rm ABS}\eta_{\rm N}}{\rho_{\rm ABS}\eta_{\rm L}} \right)^{\frac{1}{\alpha_{\rm N}}}r^{\frac{\alpha_{\rm L}}{\alpha_{\rm N}}}, &\quad \text{if } r\geq l^i_{\rm L-N}
    \end{matrix}\quad,\right. \quad\quad & d_{\rm NT}(r)= \left( \frac{\rho_{\rm T}\eta_{\rm T}}{\rho_{\rm ABS}\eta_{\rm N}} \right)^{\frac{1}{\alpha_{\rm T}}}r^{\frac{\alpha_{\rm N}}{\alpha_{\rm T}}},
\end{align}

where $l^i_{Q_1-Q_2}=\left( \frac{\rho_{Q_1}\eta_{Q_1}}{\rho_{Q_2}\eta_{Q_2}} \right)^{\frac{1}{\alpha_{Q_1}}}h_{u_i}^{\frac{\alpha_{Q_2}}{\alpha_{Q_1}}},\ \left \{ Q_1,Q_2 \right \}\subset \left \{\rm L,N,T \right \}$ and $\rho_{\rm L}=\rho_{\rm N}=\rho_{\rm ABS}$. Those variables are used to make sure that the distances to the nearest interfering base stations are not less than their altitudes which is an impossible event.

\end{nrem}

\subsection{Association probabilities}

As we already mentioned, we adopt the strongest average received power association scheme. Hence, based on the assumption that the expectation of the channel fading power gains is equal to one ($\mathbb{E}[G_{\rm N}]=\mathbb{E}[G_{\rm L}]=\mathbb{E}[G_{\rm T}]=1$), we can determine the serving base station by comparing the received powers coming from the cluster ABS, the closest LoS ABS associated with a GS located in ring $i$ for each $i\in \left[1,N\right]$, the closest NLoS ABS associated with a GS located in ring $i$ for each $i\in \left[1,N\right]$, and the closest TBS, without taking into consideration the channel fading. 
{
The typical user can be served only by one BS. In the following Lemmas, we will provide the probabilities that the typical user is associated with each type of BS.
}

\begin{lemma}\label{Lm3}
The probability that the typical user is associated with a NLoS ABS located at a distance $r$ from the typical user and associated with a GS situated in the ring $i$ is given by 

\begin{equation}\label{A_N1}
    \begin{split}
        \mathbb{A}_{\rm N_i}&(r)=\prod_{j \in \left[1,N\right], j \neq i}  \exp\left(-2\pi p_j\delta\lambda_C \int_{0}^{\sqrt{\max(r,h_{u_j})^2-h_{u_j}^2}}xp_{N_j}(x){\rm d}x\right)\\
        &\times \prod_{j \in \left[1,N\right]}  \begin{pmatrix}
        \exp\left(-2\pi p_j\delta\lambda_C \int_{0}^{\sqrt{d_{\rm NL}^2(r)-h_{u_j}^2}}xp_{L_j}(x){\rm d}x\right)
        \end{pmatrix} \times \exp\left(-\pi\lambda_{\rm T}d_{\rm NT}^2(r)\right) \mathbbm{1}_{[0,+\infty[}(r)\\
        &\times \begin{pmatrix}
        p_{\rm out}+\sum_{j \in \left[1,N\right]} p_j \int_{d_{\rm NL}(r)}^{+\infty}p_{L_j}\left( \sqrt{x^2-h_{u_j}^2} \right)f_{D_{u_j,r}}(x)\ {\rm d}x \\
         + \sum_{j \in \left[1,N\right]} p_j \int_{\max(h_{u_j},r)}^{+\infty}p_{N_j}\left( \sqrt{x^2-h_{u_j}^2} \right)f_{D_{u_j,r}}(x)\ {\rm d}x
        \end{pmatrix}.
    \end{split}
\end{equation}

{The probability that the typical user is associated with a LoS ABS located at a distance $r$ from the typical user and associated with a GS situated in the ring $i$ is given by the same expression as (\ref{A_N1}) replacing the index $N$ by $L$ and $L$ by $N$.}

\begin{IEEEproof}
See Appendix~\ref{app:Lm3}.
\end{IEEEproof}
\end{lemma}




\begin{lemma}\label{Lm2}
The probability that the typical user is associated with a TBS that is located at a distance $r$ from the typical user is given by

\begin{equation}\label{A_T}
    \begin{split}
        \mathbb{A}_{\rm T}(r)
        =& \prod_{i \in \left[1,N\right]}
        \begin{pmatrix}
        \exp\left(-2\pi p_i\delta\lambda_C \int_{0}^{\sqrt{d_{\rm TN_i}^2(r)-h_{u_i}^2}}xp_{N_i}(x){\rm d}x\right)
        \end{pmatrix}\\
        &\times \prod_{i \in \left[1,N\right]}
        \begin{pmatrix}
        \exp\left(-2\pi p_i\delta\lambda_C \int_{0}^{\sqrt{d_{\rm TL_i}^2(r)-h_{u_i}^2}}xp_{L_i}(x){\rm d}x\right)
        \end{pmatrix}\\
        &\times \begin{pmatrix}
        p_{\rm out} + \sum_{j \in \left[1,N\right]} p_j \int_{d_{\rm TN_j}(r)}^{+\infty}p_{N_j}\left( \sqrt{x^2-h_{u_j}^2} \right)f_{D_{u_j,r}}(x)\ {\rm d}x\\ 
        + \sum_{j \in \left[1,N\right]} p_j \int_{d_{\rm TL_j}(r)}^{+\infty}p_{L_j}\left( \sqrt{x^2-h_{u_j}^2} \right)f_{D_{u_j,r}}(x)\ {\rm d}x
        \end{pmatrix}.
    \end{split}
\end{equation}
where $d_{\rm TN_i}(r)$, $d_{\rm TL_i}(r)$ are given in Remark \ref{RQ}

\begin{IEEEproof}
the proof proceeds in a similar manner to Lemma \ref{Lm3}. Therefore, it is omitted here.
\end{IEEEproof}
\end{lemma}

\begin{lemma}\label{Lm5}
Given that the cluster ABS is a LoS ABS located at a distance $r$ from the typical user and associated with a GS located in the ring $j$, the probability that the typical user is associated with it  is given by

\begin{equation}\label{A_CL}
    \begin{split}
        \mathbb{A}_{\rm CL_j}(r)
        =&\prod_{i \in \left[1,N\right]}  \exp\left(-2\pi p_i\delta\lambda_C \int_{0}^{\sqrt{\max(r,h_{u_i})^2-h_{u_i}^2}}xp_{L_i}(x){\rm d}x\right) \times \exp\left(-\pi\lambda_{\rm T}d_{\rm LT}^2(r)\right)\\
        &\times \prod_{i \in \left[1,N\right]} \exp\left(-2\pi p_i\delta\lambda_C \int_{0}^{\sqrt{d_{\rm LN_i}^2(r)-h_{u_i}^2}}xp_{N_i}(x){\rm d}x\right).
    \end{split}
\end{equation}

{Given that the cluster ABS is a NLoS ABS located at a distance $r$ from the typical user and associated with a GS located in the ring $j$, the probability that the typical user is associated with it  is given by the same expression as (\ref{A_CL}) replacing the index $L$ by $N$ and $L$ by $N$.}

\begin{IEEEproof}
See Appendix~\ref{app:Lm5}.
\end{IEEEproof}
\end{lemma}



\begin{cor}\label{cor5}
the probability that the typical user is associated with the cluster ABS is given by
\begin{equation}
    \mathbb{A}_{\rm CABS}(r)= \sum_{j=1}^N p_j \left(\mathbb{A}_{\rm CN_j}(r) p_{N_j}(\sqrt{r^2-h_{u_j}^2})+\mathbb{A}_{\rm CL_j}(r) p_{L_j}(\sqrt{r^2-h_{u_j}^2})\right).
\end{equation}

\end{cor}
\subsection{Interference}

Interference is a fundamental feature of wireless communication systems, as multiple transmissions frequently occur at the same time over the same communication medium. As a result, it has the potential to significantly reduce the reliability 
of a wireless communication system.

As we already mentioned, we assume that ABSs and TBSs operate on the same frequency spectrum and thus interfere with one another. We also assume that all BSs have data to send. The interference in this case is equal to the total of all received powers from all BSs except the serving BS.  Its expression was given in (\ref{interference}).

\begin{lemma}\label{Lm7}
Supposing that the origin of the coordinate system is the typical user and denoting $z_0$ the location of the serving BS, the Laplace transform of the interference power conditioned

on the serving BS being at a distance $r$ from the typical user is given by

\begin{equation}\label{L_I}
    \begin{split}
        L_I(s)
        =& \exp\left(-\int_{a(r)}^{\infty} \left[1- \left(\frac{m_{\rm T}}{m_{\rm T}+s \rho_{\rm T}\eta_{\rm T}z^{-\alpha_{\rm T}}} \right)^{m_{\rm T}}\right]2\pi \lambda_{\rm T} z \ {\rm d}z \right)\\
        &\times \prod_{i\in\left[1,N\right]} \exp\left(-\int_{b_i(r)}^{\infty} \left[1- \left(\frac{m_{\rm L}}{m_{\rm L}+s \rho_{\rm ABS}\eta_{\rm L}\sqrt{z^2+h_{u_i}^2}^{-\alpha_{\rm L}}} \right)^{m_{\rm L}}\right]2\pi p_i \delta\lambda_C p_{L_i}(z) z\ {\rm d}z \right)\\
        &\times \prod_{i\in\left[1,N\right]} \exp\left(-\int_{c_i(r)}^{\infty} \left[1- \left(\frac{m_{\rm N}}{m_{\rm N}+s \rho_{\rm ABS}\eta_{\rm N}\sqrt{z^2+h_{u_i}^2}^{-\alpha_{\rm N}}} \right)^{m_{\rm N}}\right]2\pi p_i \delta\lambda_C p_{N_i}(z) z\ {\rm d}z \right)\\
        & \times\left(p_{\rm out}+\sum_{j\in\left[1,N\right]} \frac{p_j}{k_j(r)}
        \begin{pmatrix}
        \int^{\infty}_{e(r)} \left(\frac{m_{\rm L}}{m_{\rm L}+s \rho_{\rm ABS}\eta_{\rm L}x^{-\alpha_{\rm L}}} \right)^{m_{\rm L}}p_{L_j}(\sqrt{x^2-h_{u_j}^2}) f_{D_{u_j,r}}(x)\ {\rm d}x \\ 
        + \int^{\infty}_{l(r)}\left(\frac{m_{\rm N}}{m_{\rm N}+s \rho_{\rm ABS}\eta_{\rm N}x^{-\alpha_{\rm N}}} \right)^{m_{\rm N}}p_{N_j}(\sqrt{x^2-h_{u_j}^2}) f_{D_{u_j,r}}(x)\ {\rm d}x
        \end{pmatrix}\right),\\
    \end{split}
\end{equation}
where
\begin{equation}
    \left\{\begin{matrix}
    a(r)=r;\ b_i(r)=\sqrt{d^2_{\rm TL_i}(r)-h^2_{u_i}};\ c_i(r)=\sqrt{d^2_{\rm TN_i}(r)-h^2_{u_i}}\ ,\text{if }z_0 \in \Phi_{\rm T} , \\ 
    a(r)=d_{\rm LT}(r);\ b_i(r)=\sqrt{\max(r^2-h^2_{u_i},0)};\ c_i(r)=\sqrt{d^2_{\rm LN_i}(r)-h^2_{u_i}}\ ,\text{if }z_0 \in \Phi_{\rm L_j},\\ 
    a(r)=d_{\rm NT}(r);\ b_i(r)=\sqrt{d^2_{\rm NL}(r)-h^2_{u_i}};\ c_i(r)=\sqrt{\max(r^2-h^2_{u_i},0)}\ ,\text{if }z_0 \in \Phi_{\rm N_j},
    \end{matrix}\right.
\end{equation}  
\begin{equation}
    \left\{\begin{matrix}
    \begin{aligned}
    e(r)= d_{\rm TL_j}(r); l(r)=d_{\rm TN_j}(r); k_j(r)=\mathbb{P}(P^r_{\rm T}>P^r_{\rm CABS_j}|D_{\rm T}=r) ,\ &\text{if }z_0 \in \Phi_{\rm T}, \\
    e(r)= \max\left(r,h_{u_j}\right); l(r)=d_{\rm LN_j}(r) ; k_j(r)=\mathbb{P}(P^r_{\rm L_i}>P^r_{\rm CABS_j}|D_{\rm L_i}=r),\ &\text{if }z_0 \in\Phi_{\rm L_i},\forall i \in \left[1,N\right],\\ 
    e(r)= d_{\rm NL}(r); l(r)=\max\left(r,h_{u_j}\right); k_j(r)=\mathbb{P}(P^r_{\rm N_i}>P^r_{\rm CABS_j}|D_{\rm N_i}=r) ,\ &\text{if }z_0 \in\Phi_{\rm N_i}, \forall i \in \left[1,N\right],
    \end{aligned}
    \end{matrix}\right.
\end{equation}


Moreover, if the serving base station is the cluster ABS, the Laplace transform of the interference power is defined as

\begin{equation}\label{L_IC}
    \begin{split}
        L_{I}(s)=& \exp\left(-\int_{a(r)}^{\infty} \left[1- \left(\frac{m_{\rm T}}{m_{\rm T}+s \rho_{\rm T}\eta_{\rm T}z^{-\alpha_{\rm T}}} \right)^{m_{\rm T}}\right]2\pi \lambda_{\rm T} z \ {\rm d}z \right) \\
        \times& \prod_{i\in\left[1,N\right]} \exp\left(-\int_{b_i(r)}^{\infty} \left[1- \left(\frac{m_{\rm L}}{m_{\rm L}+s \rho_{\rm ABS}\eta_{\rm L}\sqrt{z^2+h_{u_i}^2}^{-\alpha_{\rm L}}} \right)^{m_{\rm L}}\right]2\pi p_i \delta\lambda_C p_{L_i}(z) z\ {\rm d}z \right)\\
        \times& \prod_{i\in\left[1,N\right]} \exp\left(-\int_{c_i(r)}^{\infty} \left[1- \left(\frac{m_{\rm N}}{m_{\rm N}+s \rho_{\rm ABS}\eta_{\rm N}\sqrt{z^2+h_{u_i}^2}^{-\alpha_{\rm N}}} \right)^{m_{\rm N}}\right]2\pi p_i \delta\lambda_C p_{N_i}(z) z\ {\rm d}z \right),
    \end{split}
\end{equation}
where
\begin{equation}
    \left\{\begin{matrix}
    a(r)=d_{\rm LT}(r);\ b_i(r)=\sqrt{\max(r^2-h^2_{u_i},0)};\ c_i(r)=\sqrt{d^2_{\rm LN_i}(r)-h^2_{u_i}}, \text{ \small if the ABS is  LoS }, \\ 
    a(r)=d_{\rm NT}(r);\ b_i(r)=\sqrt{d^2_{\rm NL}(r)-h^2_{u_i}};\ c_i(r)=\sqrt{\max(r-h^2_{u_i},0)} ,\text{ \small if the ABS is NLoS}.
    \end{matrix}\right.
\end{equation}  

\begin{IEEEproof}
See Appendix~\ref{app:Lm7}.
\end{IEEEproof}

\end{lemma}

\subsection{Exact coverage probability}

{\color{black}

The coverage probability is one of the key performance quantities. It is generally defined as the probability that the SINR is greater than a designated threshold $\gamma$
\begin{equation}
    P_{\rm cov}=\mathbb{P}(\rm SINR>\gamma).
\end{equation}

\begin{theorem}\label{thm2}
The overall coverage probability of the proposed system is given by
\begin{equation}
    \begin{split}
        P_{\rm cov}
        =& \sum_{i\in\left[1,N\right]} \int_{h_{u_i}}^{\infty} P^{\rm L_i}_{\rm cond}(r) A_{\rm L_i}(r) f_{D_{\rm L_i}}(r) \ {\rm d} r + \sum_{i\in\left[1,N\right]} \int_{h_{u_i}}^{\infty} P^{\rm N_i}_{\rm cond}(r) A_{\rm N_i}(r) f_{D_{\rm N_i}}(r) \ {\rm d} r \\
        &+ \sum_{j\in\left[1,N\right]} p_j \int_{h_{u_j}}^{\infty} P^{\rm CL_j}_{\rm cond}(r) A_{\rm CL_j}(r)+P^{\rm CN_j}_{\rm cond}(r) A_{\rm CN_j}(r) f_{D_{u_j,r}} (r) \ {\rm d}r \\
        &+\int_0^{\infty} P^{\rm T}_{\rm cond}(r) A_{\rm T}(r) f_{D_T}(r) \ {\rm d}r 
    \end{split}
\end{equation}    
where $P^{\{L_i,N_i,CL_j,CN_j,T\}}_{\rm cond}$ is the conditional coverage probability given that the typical user is associated with: a LoS ABS linked to a GS located in the ring $i$, a NLoS ABS linked to a GS located in the ring $i$, the cluster ABS which is a LoS ABS linked to a GS in the ring $j$, the cluster ABS which is a NLoS ABS linked to a GS in the ring $j$, or the TBS.

The conditional coverage probabilities are given by
\begin{equation} \label{40}
   P^{\rm Q'}_{\rm cond}(r)=\sum_{k=0}^{m_{\rm Q}-1} \left[\frac{(-s)^k}{k!}\frac{\partial^k}{\partial s^k}L_U(s) \right]_{s=\mu_{\rm Q}(r)}, 
\end{equation}
where $\mu_Q(r)=m_Q\gamma (P_Q\eta_Q)^{-1} r^{\alpha_Q},\ U=I+\sigma^2$ and $(Q',Q)= \{\rm (T,T),(L_i,L),(N_i,N),\rm (CL_j,L),$ $( CN_j,N) \}$. Clearly, $U$ depends on the type of the serving BS. Also $ A_{\rm N_i}, A_{\rm L_i}, A_{\rm T}, A_{\rm CL_j}$ and $A_{\rm CN_j}$ are given respectively in Lemma \ref{Lm3}, 
, Lemma \ref{Lm2} and Lemma \ref{Lm5}.

\begin{IEEEproof}
See Appendix~\ref{app:thm2}.
\end{IEEEproof}

\end{theorem}

}

\subsection{Approximate coverage probability}

As can be seen in (\ref{40}), The evaluation of conditional coverage probabilities necessitates the evaluation of higher order derivatives of the Laplace transform. The larger the shape parameters $m_{\rm N}, m_{\rm L}$, and $m_{\rm T}$ are, the more complicated the computation becomes.
Therefore, following the same approximations as \cite{Alzenad},  we use a tight bound of the CDF of the Gamma distribution to provide an approximate evaluation of the coverage probability.
{\color{black}
\begin{theorem}\label{thm3}
Using the upper bound of the 
CDF of the Gamma distribution, the overall coverage probability can be approximated as 
\begin{equation}
    \begin{split}
        \Tilde{P}_{\rm cov}
        =& \sum_{i\in\left[1,N\right]} \int_{h_{u_i}}^{\infty} \Tilde{P}^{\rm L_i}_{\rm cond}(r) A_{\rm L_i}(r) f_{D_{\rm L_i}}(r) \ {\rm d} r + \sum_{i\in\left[1,N\right]} \int_{h_{u_i}}^{\infty} \Tilde{P}^{\rm N_i}_{\rm cond}(r) A_{\rm N_i}(r) f_{D_{\rm N_i}}(r) \ {\rm d} r \\
        &+ \sum_{j\in\left[1,N\right]} p_j \int_{h_{u_j}}^{\infty} \Tilde{P}^{\rm CL_j}_{\rm cond}(r) A_{\rm CL_j}(r)+\Tilde{P}^{\rm CN_j}_{\rm cond}(r) A_{\rm CN_j}(r) f_{D_{u_j,r}} (r) \ {\rm d}r \\
        &+\int_0^{\infty} \Tilde{P}^{\rm T}_{\rm cond}(r) A_{\rm T}(r) f_{D_T}(r) \ {\rm d}r 
    \end{split}
\end{equation}    
where $\Tilde{P}^{\rm L_i}_{\rm cond}, \Tilde{P}^{\rm N_i}_{\rm cond}, \Tilde{P}^{\rm CL_j}_{\rm cond}, \Tilde{P}^{\rm CN_j}_{\rm cond}$ and $\Tilde{P}^{\rm T}_{\rm cond}$ are the approximate conditional coverage probabilities, and are given by

\begin{equation}
    \Tilde{P}^{\rm {Q}'}_{\rm cond}(r)=\sum_{k=1}^{m_{Q}}\binom{m_{Q}}{k} (-1)^{k+1} L_U \left(k\beta_{Q}\mu_{Q}(r)\right),
\end{equation}
where $\beta_{Q}=(m_Q!)^{\frac{-1}{m_Q}}$, $\mu_Q(r)=m_Q\gamma (P_Q\eta_Q)^{-1} r^{\alpha_Q}$ and $(Q',Q)= \{\rm (T,T),(L_i,L),(N_i,N),$ $\rm (CL_j,L),( CN_j,N) \}$.

\begin{IEEEproof}
See Appendix~\ref{app:thm3}.
\end{IEEEproof}

\end{theorem}
}

\section{Numerical results}

In this section, we validate the analytical expression of the coverage probability with Monte Carlo simulations and we discuss the impact of various system parameters on the coverage performance.
\begin{table}[h]
\centering
\caption{Simulation parameters {\cite{optDep,Alzenad,9420290,3-Dplacement}}.}\label{tab}
\begin{tabular}{|c|c|c|c|} 
\hline
\multirow{2}{*}{Parameter} & \multirow{2}{*}{symbol} & \multicolumn{2}{l|}{~ ~ ~ ~ ~  value }  \\ 
\cline{3-4}
                    &        & Urban     & Suburban                 \\ 
\hhline{|====|}
\multicolumn{1}{|c|}{Cluster raduis}&$R_0$          & \multicolumn{2}{c|}{$200$m} \\ 
\hline
\multicolumn{1}{|c|}{Number of rings}&$N$          & \multicolumn{2}{c|}{50} \\ 
\hline
\multicolumn{1}{|c|}{Maximum tether length}&$T_{\rm max}$          & \multicolumn{2}{c|}{$80$m \cite{3-Dplacement}} \\ 
\hline

\multicolumn{1}{|c|}{Minimum inclination angle} &$\theta_{\rm min}$                & $15.3^{\circ}$ \cite{3-Dplacement}     & $10.6^{\circ}$   \cite{3-Dplacement}        \\ 
\hline
\multicolumn{1}{|c|}{LoS probability parameter}&$a$                         & 13 \cite{optDep}       & 4.88 \cite{9420290}                     \\ 
\hline
\multicolumn{1}{|c|}{LoS probability parameter}&$b$                         & 0.21 \cite{optDep}     & 0.429   \cite{9420290}                  \\ 
\hline
\multicolumn{1}{|c|}{Altitude of buildings}&$h_n$                       & $15$m  \cite{3-Dplacement}       & $8$m  \cite{3-Dplacement}                      \\ 
\hline
\multicolumn{1}{|c|}{Density of buildings}&$\lambda_b$                 & 500/$\rm Km^2$  \cite{3-Dplacement} & 750/$\rm Km^2$   \cite{3-Dplacement}                  \\ 
\hline
\multicolumn{1}{|c|}{accessibility factor}&$\kappa_b$          & \multicolumn{2}{c|}{0.02} \\ 
\hline
\multicolumn{1}{|c|}{LoS attenuation coefficient}&$\eta_{\rm L}$           & 0.4   \cite{optDep}     & 0.9772   \cite{9420290}                \\ 
\hline
\multicolumn{1}{|c|}{NLoS attenuation coefficient}&$\eta_{\rm N}$           & 0.005  \cite{optDep}    & 0.0079   \cite{9420290}                \\ 
\hline
\multicolumn{1}{|c|}{TBS attenuation coefficient}&$\eta_{\rm T}$            & 0.1   \cite{optDep}     & 0.69      \cite{9420290}               \\ 
\hline
\multicolumn{1}{|c|}{TBS density}&$\lambda_{\rm T}$          & 10/$\rm Km^2$ & 1.5/$\rm Km^2$ \\ 
\hline
\multicolumn{1}{|c|}{ The cluster center density}&$\lambda_{C}$          & 20/$\rm Km^2$ & 5/$\rm Km^2$ \\ 
\hline
Noise power&$\sigma^2$                  & $10^{-8} W$& $10^{-12} W$              \\
\hline
\multicolumn{1}{|c|}{LoS pathloss exponent}&$\alpha_{\rm L}$         & \multicolumn{2}{c|}{2 \cite{Alzenad}}  \\ 
\hline
\multicolumn{1}{|c|}{NLoS pathloss exponent}&$\alpha_{\rm N}$         & \multicolumn{2}{c|}{3 \cite{Alzenad}}                         \\ 
\hline
\multicolumn{1}{|c|}{TBS pathloss exponent}&$\alpha_{\rm T}$          & \multicolumn{2}{c|}{3 \cite{Alzenad}} \\ 
\hline
\multicolumn{1}{|c|}{ABS transmit power}&$\rho_{\rm ABS}$               & \multicolumn{2}{c|}{$1 W$ \cite{Alzenad}}  \\ 
\hline
TBS transmit power&$\rho_{\rm T}$               & \multicolumn{2}{c|}{$10 W$ \cite{Alzenad}}  \\ 
\hline
\multicolumn{1}{|c|}{LoS Nakagami-m fading term}&$m_{\rm L}$         & \multicolumn{2}{c|}{2 \cite{Alzenad}}  \\ 
\hline
\multicolumn{1}{|c|}{NLoS Nakagami-m fading term}&$m_{\rm N}$         & \multicolumn{2}{c|}{1 \cite{Alzenad}}                         \\ 
\hline
\multicolumn{1}{|c|}{TBS Nakagami-m fading term}&$m_{\rm T}$          & \multicolumn{2}{c|}{1 \cite{Alzenad}} \\ 
\hline

\multicolumn{1}{|c|}{SINR threshold}&$\gamma$          & \multicolumn{2}{c|}{1} \\ 
\hline
\end{tabular}
\end{table}
First, we numerically solve the optimization problem (\ref{optpb}) to determine the optimal values that minimize the average pathloss. 
Then deploying T-UAVs in their optimal placements, we use Matlab to evaluate the main expression which is the coverage probability expression and run 10000 Monte Carlo iterations. We consider a square region that measures $400$Km on each side. In each iteration, we generate the ABS PPPs $\left \{ \Phi_{\rm LABS_i},\Phi_{\rm NABS_i}, i \in \left [ 1,N \right ] \right \}$, the TBS PPP $\Phi_{\rm TBS}$, and the coordinates of the typical user which is uniformly distributed inside clusters. Then, we calculate the received powers at the typical user from each type of BS and compare them to determine the serving BS. Based on the type of the serving BS, we calculate the interference and the SINR. Finally, comparing the SINR to the chosen threshold, we deduce the coverage probability. Throughout this section and unless otherwise specified, we use the default system parameters listed in Table \ref{tab}, which mostly correspond to the parameters listed in \cite{optDep,Alzenad,9420290}.

\begin{figure}
\centering
\begin{minipage}[c]{.48\textwidth}
  \centering
  \includegraphics[width=.97\linewidth]{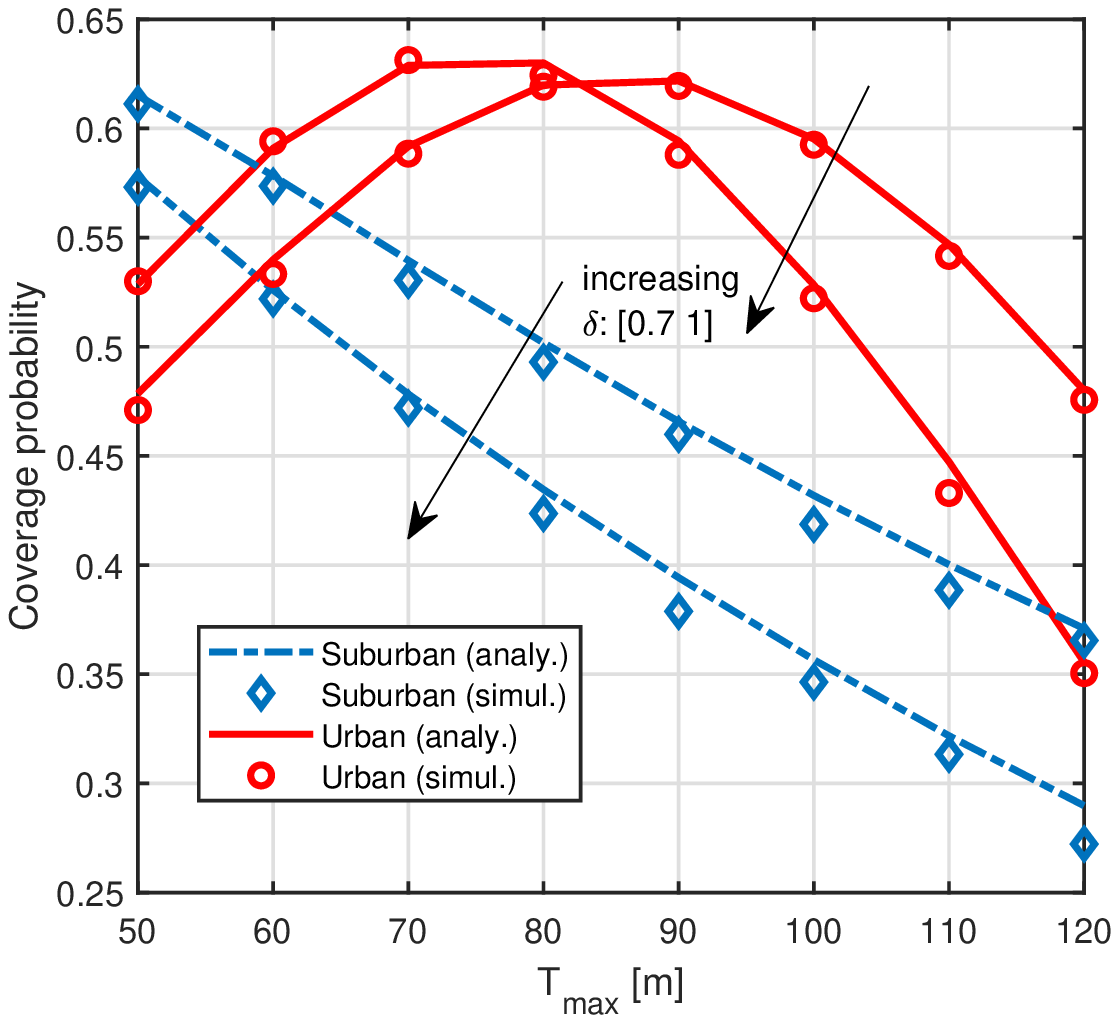}
  \captionof{figure}{Coverage probability for different values of $T_{\rm max}$.}
  \label{covT}
\end{minipage}\hfill
\begin{minipage}[c]{.48\textwidth}
  \centering
  \includegraphics[width=.95\linewidth]{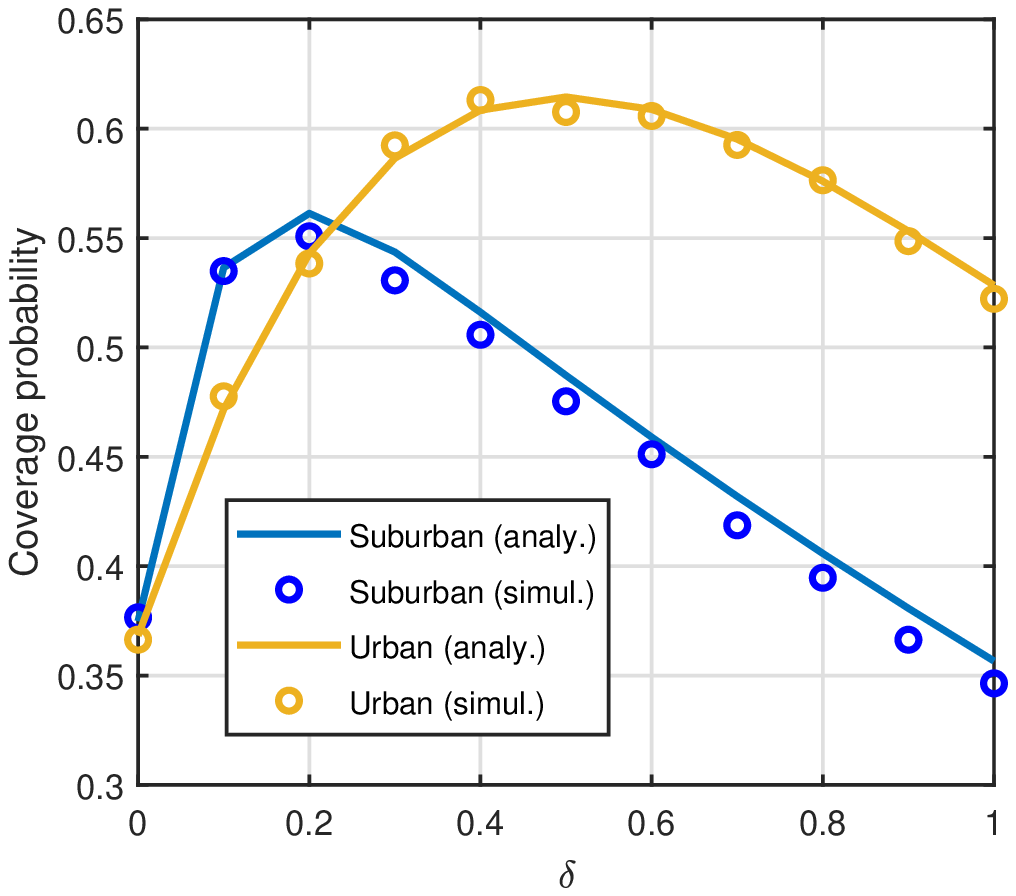}
  \captionof{figure}{Coverage probability for different values of $\delta$.}
  \label{covdelta}
\end{minipage}
\end{figure}


Fig.\ref{covT} shows the variation of coverage probability with the maximum value of the tether length $T_{\rm max}$ for different values of $\delta$ in urban and suburban environments. {Looking at Fig.3, we can see that the analytical and simulation results match with a small gap that is mostly visible in the suburban curves. This difference is due to the approximation in Theorem 3 that is  used to generate the numerical values of the coverage probability.} In suburban environment,  the coverage probability is a strictly decreasing function. It falls from 0.61 to 0.36 for $\delta=0.7$, and from 0.57 to 0.28 for  $\delta=1$. So, decreasing $\delta$ improves the coverage performance. This is explained by the fact that decreasing $\delta$ reduces the number of deployed T-UAVs, resulting in less interference and an increase in the coverage probability. In urban environment, the coverage probability increases as $T_{\rm max}$ increases.  It reaches its maximum value, then decreases. Hence, we have an optimal value of $T_{\rm max}$. For $\delta=1$, the optimal value of $T_{\rm max}$  is equal to $80m$, while it is equal to $90m$ for $\delta=0.7$. With the augmentation of $T_{\rm max}$, T-UAVs altitude increases. Thus, there are more T-UAVs experience LoS conditions with the users. This implies that the users are more likely to be associated with LoS ABS while the interference power is dominated by NLoS ABSs and TBSs. Since the LoS transmissions have the best channel conditions, the coverage probability will improve. However, as the maximum tether length is increased further, we have more interfering LoS ABS. Thus, the interference will significantly rise, causing a coverage probability drop.   


Furthermore, we notice that the coverage probability varies with $\delta$. Thus, we plot the coverage probability as a function of $\delta$, 
as illustrated in Fig.\ref{covdelta}. In Fig.\ref{covdelta}, we can see that $\delta$ has an optimal value for both urban and suburban environments. This optimal value is equal to 0.2 for suburban environment and 0.5 for urban environment. 
{\color{black} It is more likely to establish a LoS link in a suburban environment than in an urban environment due to the density and height of the surrounding buildings. Thus the number of LoS ABSs in a suburban environment is more important than in an urban environment. With the increase of the maximum tether length, the number of LoS ABSs will increase furthermore. Thus the interference will be much more important in a suburban environment which explains the fact that the coverage performance of ABS is better in an urban environment. In a suburban environment, we need to reduce the number of deployed T-UAVs to restrain the bad effect of interference and have good coverage performance.   
}

\begin{figure}
\centering
\begin{minipage}[c]{.48\textwidth}
  \centering
  \includegraphics[width=1\linewidth]{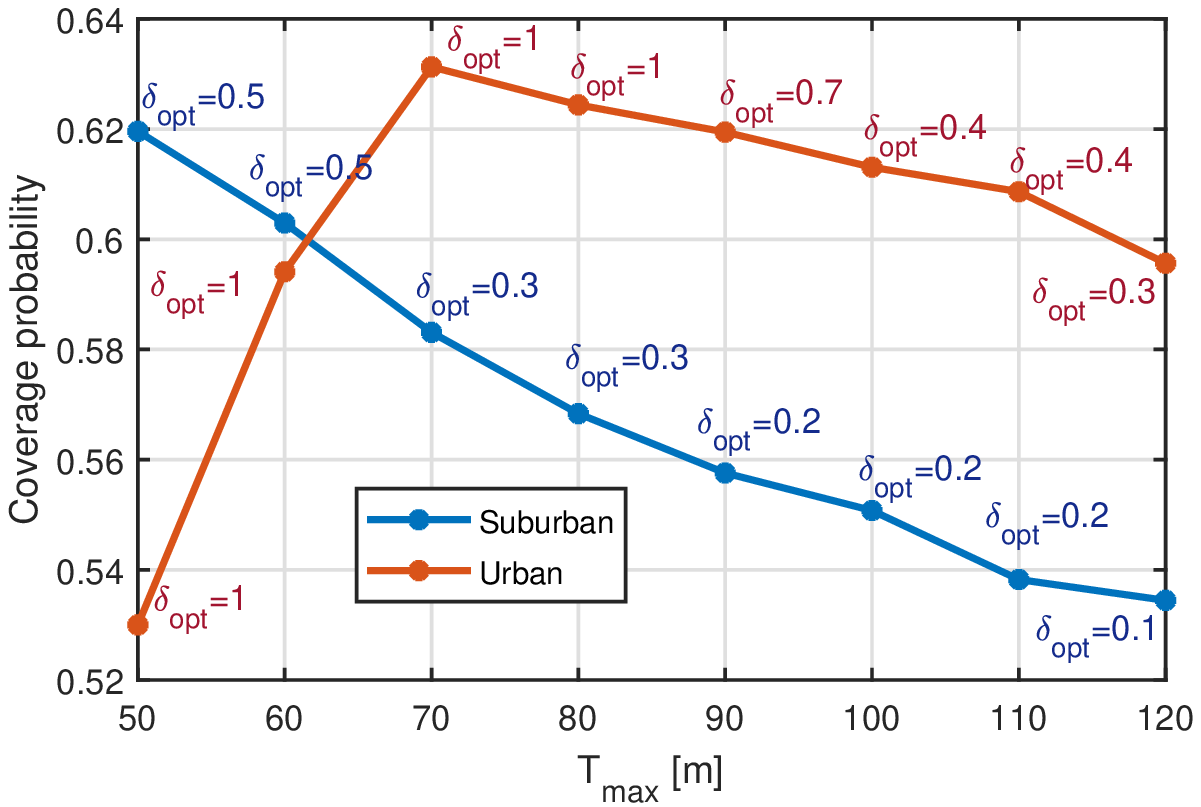}
  \captionof{figure}{Coverage probability for different values of $T_{\rm max}$ and optimal values of $\delta$.}
  \label{covTD}
\end{minipage}\hfill
\begin{minipage}[c]{.48\textwidth}
  \centering
  \includegraphics[width=1\linewidth]{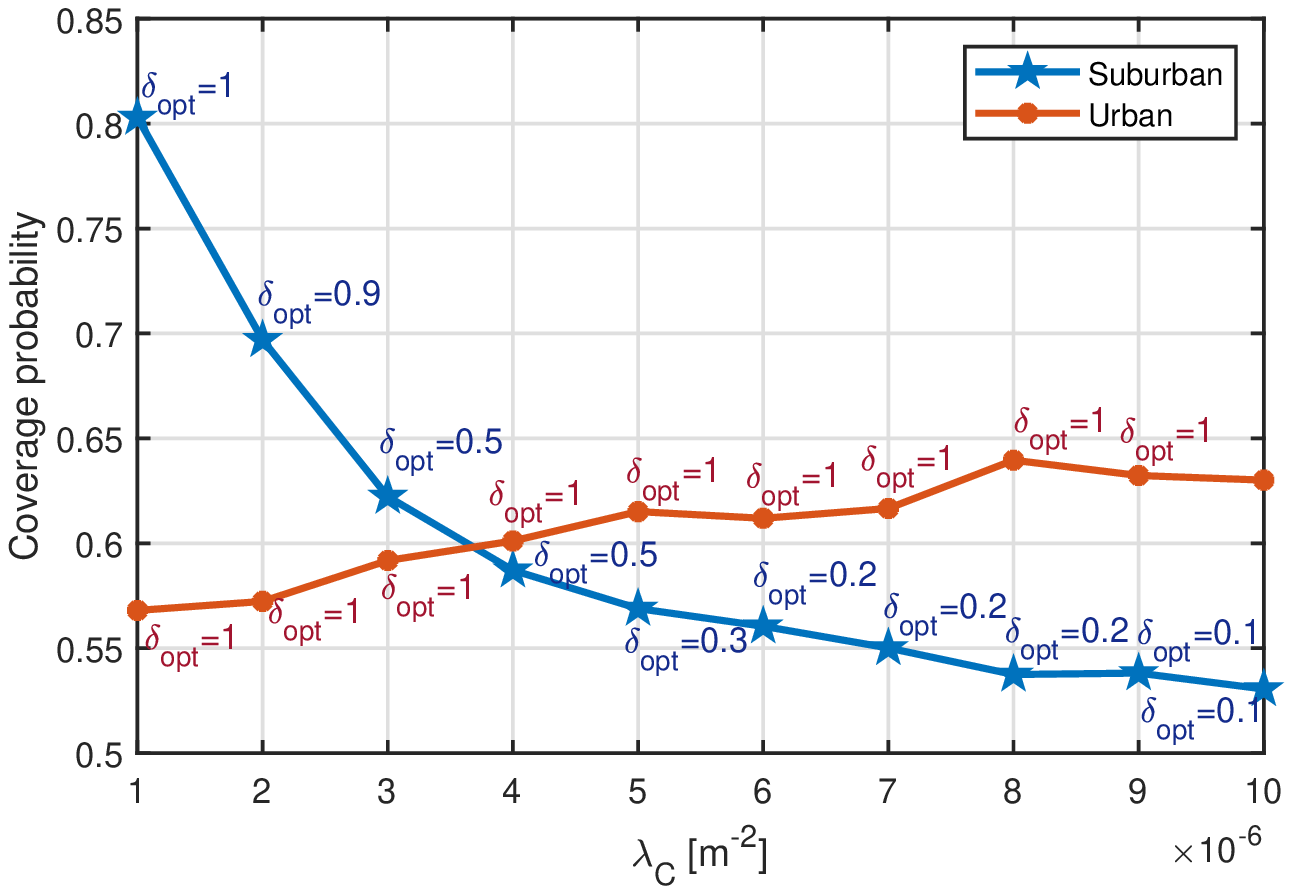}
  \captionof{figure}{Coverage probability for different values of $\lambda_C$ and optimal values of $\delta$.}
  \label{covC}
\end{minipage}
\end{figure}
Knowing that there are optimal values of $\delta$, we generate Fig.\ref{covTD} which depicts the variation of coverage probability with $T_{\rm max}$, taking into consideration the optimal values of $\delta$. The optimal value of $\delta$ decreases as $T_{\rm max}$ increases.
With the augmentation of $T_{\rm max}$, T-UAVs altitude increases. Consequently, more T-UAVs are encountering LoS conditions with the users. This increases the interference and as a result, reduces the coverage probability. So, the optimal value of $\delta$ will decrease in order to reduce the number of LoS ABS and thus the interference. Moreover, in suburban environment, the coverage probability reaches its maximum value for $T_{\rm max}=50$m and $\delta=0.5$, whereas in urban environment, it reaches its maximum value for $T_{\rm max}=70$m and $\delta=1$.

\begin{figure}[t]
\centering
\begin{minipage}[c]{.48\textwidth}  \includegraphics[width=1\linewidth]{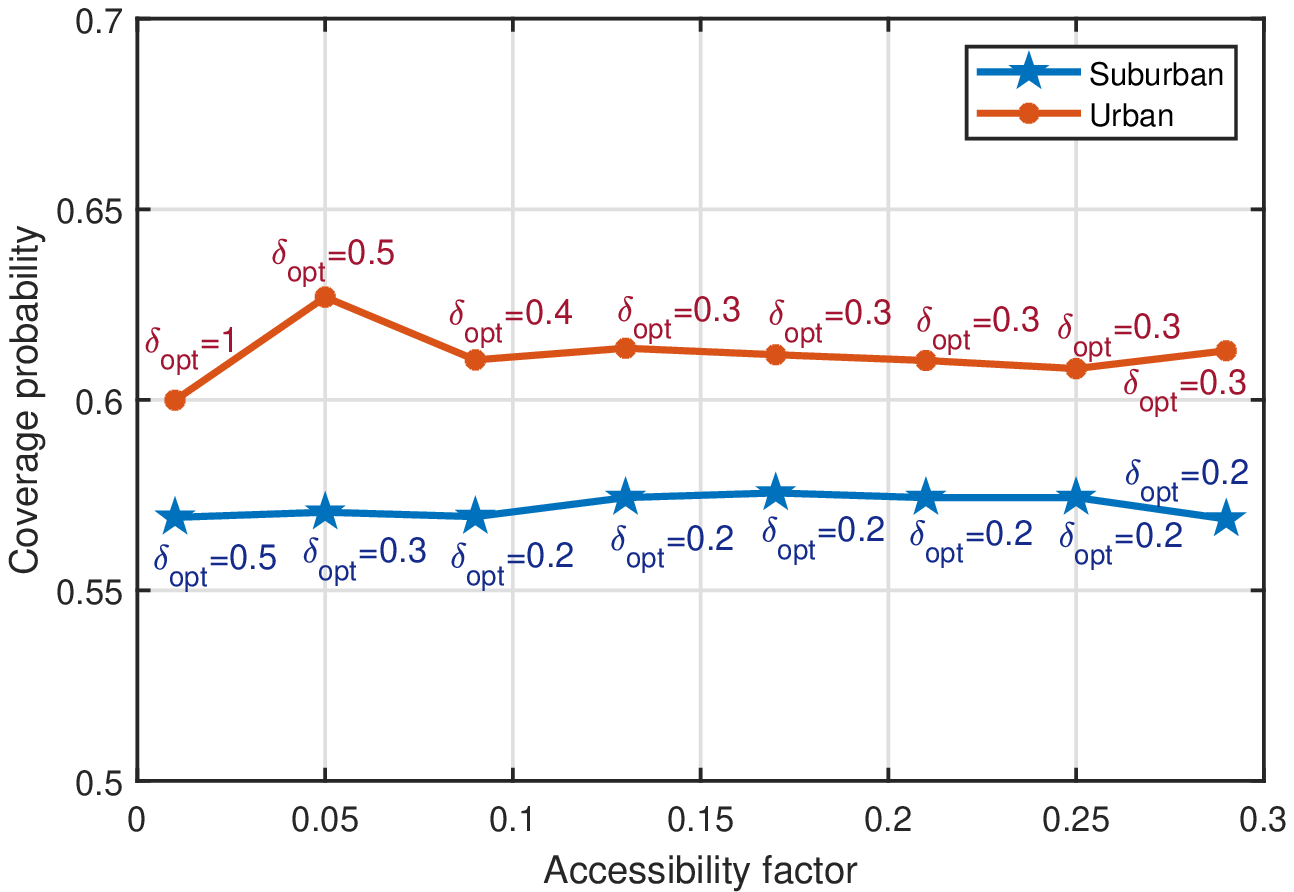}
  \captionof{figure}{Coverage probability for different values of accessibility factor and optimal values of $\delta$.}
  \label{covA}
\end{minipage}\hfill
\begin{minipage}[c]{.48\textwidth}
  \centering
  \includegraphics[width=.9\linewidth]{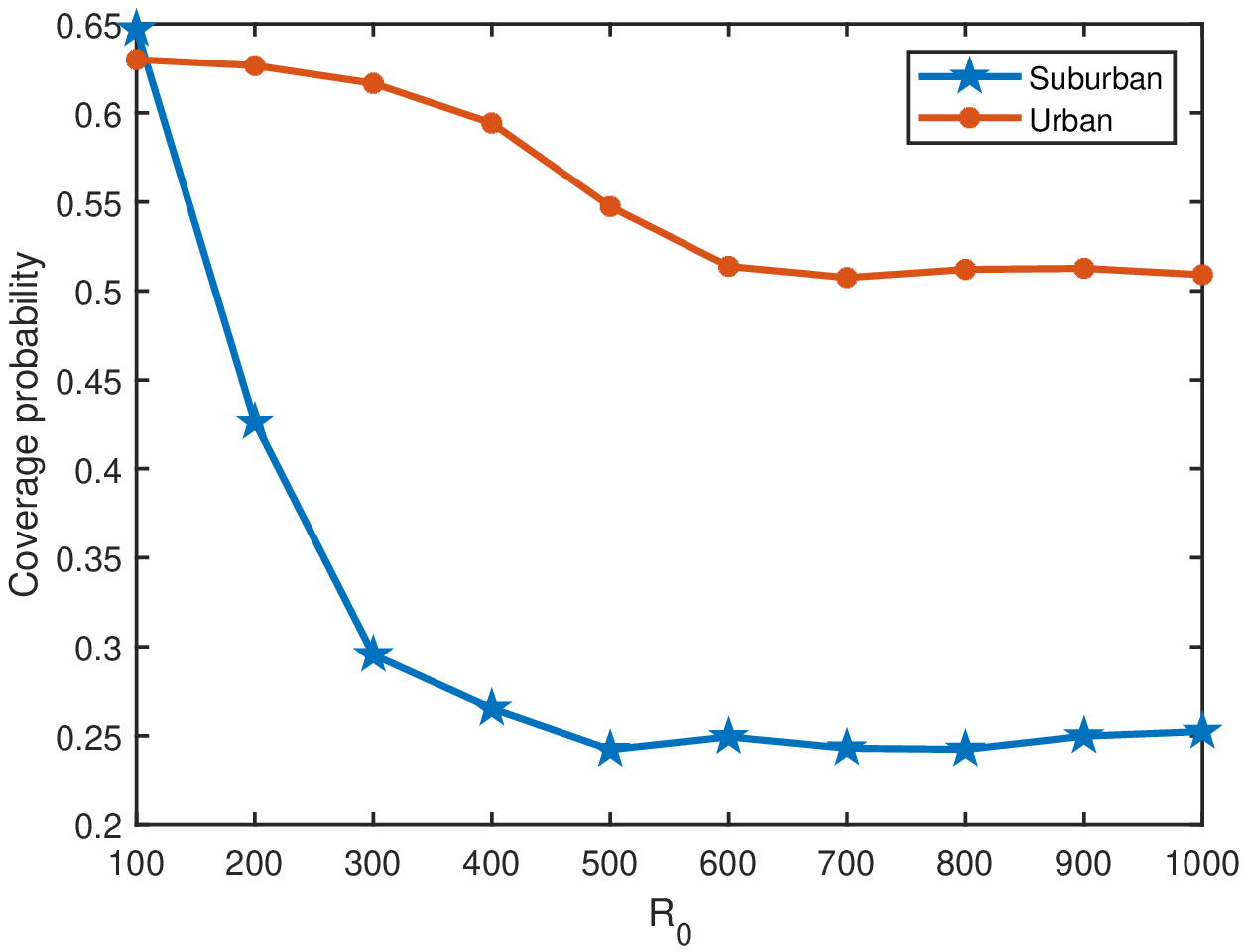}
  \captionof{figure}{Coverage probability as a function of the cluster radius.}
  \label{cov_R0}
\end{minipage}
\end{figure}

Fig.\ref{covC} depicts how the coverage probability varies with the density of the cluster centers ($\lambda_C$), taking into account the optimal values of $\delta$. In urban environments, the coverage probability increase slightly with the rise of $\lambda_C$, while the optimal value of $\delta$ remain constant and equal to 1. In suburban case, the coverage probability decreases dramatically from 0.74 to 0.22, and so does the optimal value of $\delta$ which decrease from 1 to 0.1. This can be explained by the high probability of having LoS transmission in suburban areas. For low densities ($1 \rm Km^{-2}$, $2 \rm Km^{-2}$), the typical user is more likely to be associated with LoS ABS which clarifies the high coverage probability. Increasing $\lambda_C$, the number of interfering LoS ABSs substantially rise, so does the interference which demonstrates the coverage probability drop. The optimal value of $\delta$ decreases to reduce the number of ABSs and more specificly the number of LoS ABSs.
Fig.\ref{covA} illustrates the variation of the coverage probability with the accessibility factor, taking into account the optimal values of $\delta$. We notice that, for both environments, the coverage probability is almost constant. Moreover, $\delta$ decreases from $0.5$ to $0.2$ then remain constant, in a suburban environment. While it decreases from $1$ to $0.3$ then remain constant, in an urban environment.

{\color{black}
Fig.\ref{cov_R0}  illustrates the variation of the coverage probability as a function of the cluster radius $R_0$. In an urban environment, the coverage probability slightly declines with the rise of $R_0$ and then remains steady. However, in a suburban environment, the coverage probability dramatically decreases with the increase of $R_0$. It drops from 0.65 when $R_0=100$m to 0.25 when $R_0=500$m.   The association scheme we use can help to understand this. Recall that the serving BS is the BS that delivers the greater average power to the user. Therefore, the user can be served by the cluster ABS, another ABS, or a TBS. When we increase the cluster radius, the user becomes further away from the cluster UAV. Thus, the association probability with the cluster ABS decreases and the performance of the system becomes more like a PPP, than a Poisson Cluster Point (PCP). In an urban environment, the density of ABS is high. So, even if the probability of being associated with the cluster ABS is low, the probability that the user is associated with a nearby ABS other than the cluster ABS is high. So, the coverage performance will not be affected much. While in a suburban environment, the density of ABS is relatively small. Since the system behaves like a PPP as we increase the cluster radius, the coverage probability will decrease. }

\begin{figure}[t]
\centering
\begin{minipage}[c]{.48\textwidth}  \includegraphics[width=1\linewidth]{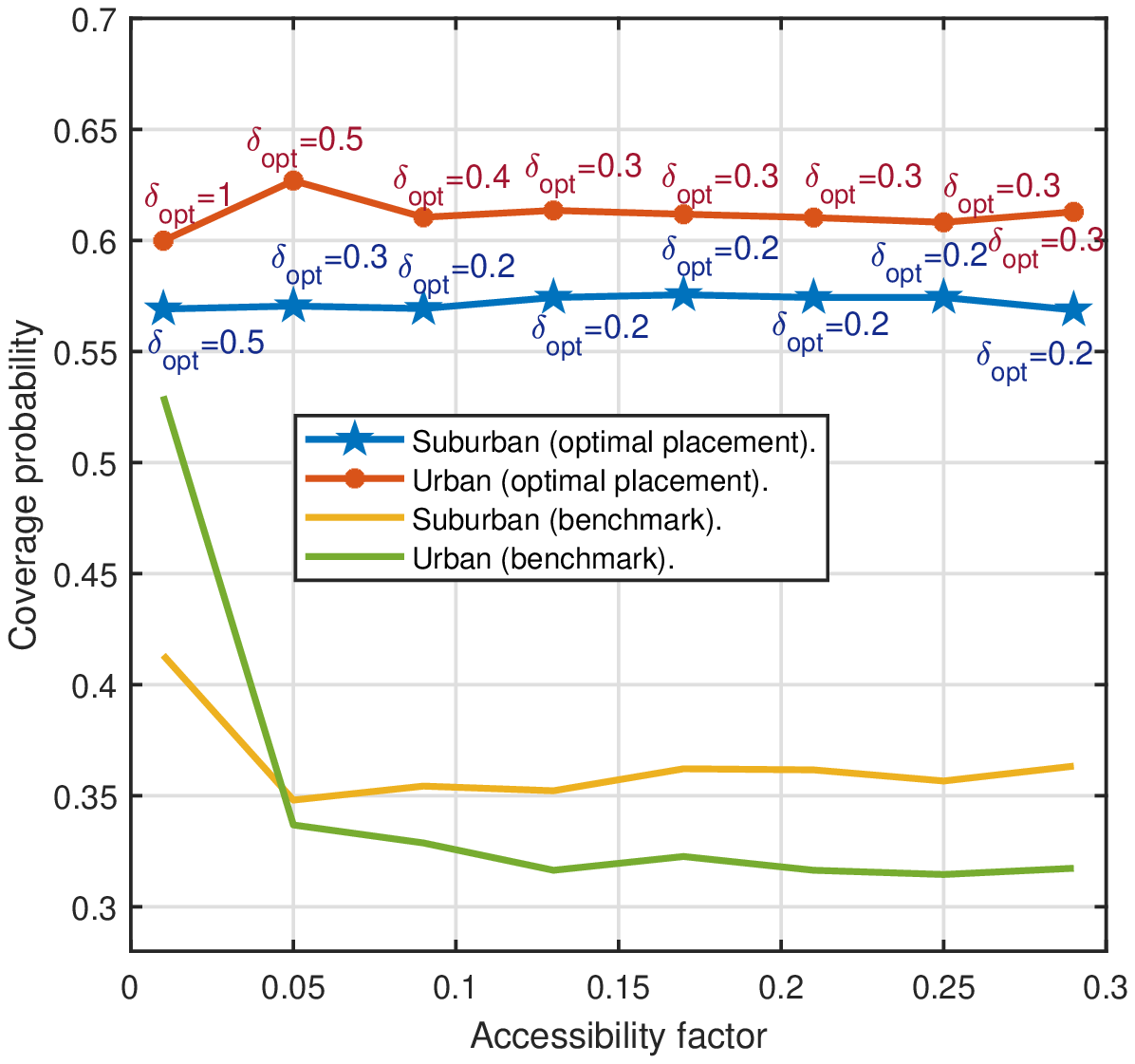}
  \captionof{figure}{Coverage probability with different values of accessibility factor for proposed and reference systems.}
  \label{Bench}
\end{minipage}\hfill
\begin{minipage}[c]{.48\textwidth}
  \centering
  \includegraphics[width=1\linewidth]{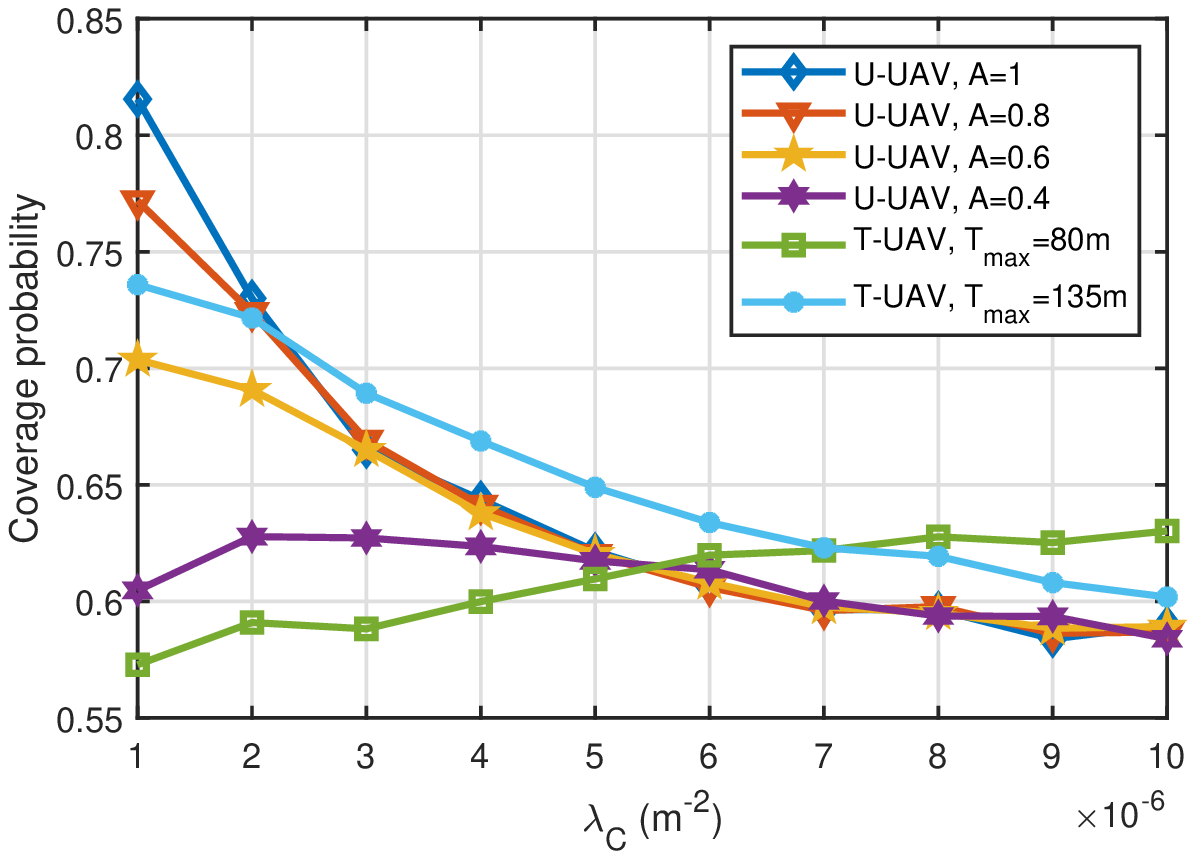}
  \captionof{figure}{Coverage probability as a function of the cluster density for different values of availability factor A.}
  \label{9}
\end{minipage}
\end{figure}

After analysing the impact of each parameter, we compare our setup with a reference one to highlight the efficiency of the proposed system. As a benchmark, we consider the same proposed setup except that the T-UAV will be deployed exactly above the GS  in each one of the clusters (inclination angle equal to $90^{\circ}$ and $\delta$ equal to 1). 
Looking at Fig.\ref{Bench}, we can clearly see that the proposed system outperforms the reference system in both environments. As a result, optimizing T-UAV placement and determining the optimal number of T-UAVs to be deployed, significantly improves the coverage performance.

Furthermore, we compare our setup to a similar setup of U-UAVs in order to highlight the differences between the two types of UAVs.
We set the maximum altitude of the U-UAV to $150m$. As illustrated in Fig.\ref{9}, we plot the coverage probability as a function of the cluster density $\lambda_C$ for different values of duty-cycle parameter A and different values of the maximum tether length $T_{\rm max}$. Those plots are generated for optimal values of delta and for urban environment. In Fig.\ref{9}, We can clearly see that, starting from $\lambda_C=2/\rm Km^2$, T-UAV with $T_{\rm max}=135m$ outperform U-UAV regardless of $A$ value, while T-UAV with $T_{\rm max}=80m$ outperform U-UAV starting from $\lambda_C=6/\rm Km^2$. T-UAV is even better than U-UAV with $A=1$ which is an unrealistic case since U-UAV can not be always available. 
Moreover, the coverage probabilities of U-UAVs with different values of $A$ decrease as $\lambda_C$ decreases and they all converge to the same value 0.58 for $\lambda_C=10/\rm Km^2$ .


\section{Conclusion}

In this paper, we have used stochastic geometry to model Tethered UAV-assisted Network where hotspots were modelled as clusters that were divided into frames to make the system analysis tractable. As a first step in the analysis, we formalized an optimization problem to determine the optimal T-UAV placement that minimizes the average pathloss. Then, deploying T-UAVs in their optimal location, we analysed the coverage performance of the system. First, we derived the distributions of the distances between a typical user and the closest BS from each type. Second, we established the association probability expressions and developed the Laplace transform of interference formulas. Finally, we proposed an expression for the coverage probability. 
Moreover, we performed Monte Carlo simulations to validate the proposed expressions and study the influence of some parameters on the coverage performance.  We deduced that the maximum tether length has an optimal value in urban environment. Hence, providing much freedom to the tether will badly affect the coverage probability because even if we reduce the average pathloss, the interference will increase. {We also provided a comparison between the proposed set-up and benchmarks to highlight the efficiency of T-UAV system.}

This work has many possible extensions. Since we focused on a stochastic geometry setup, one possible extension would be optimizing the T-UAV locations for a deterministic setup composed of a finite number of clusters with finite options for the rooftops. Another possible extension would  be to find the optimal locations of the tethered UAVs based on other metrics such as coverage probability or data rate.

\appendices

\section{Proof of Lemma~\ref{thm1}}\label{app:thm1}

We will first compute the horizontal distance between the typical user and the cluster ABS. Supposing that the GS associated with the cluster ABS is located in the ring $j$, this distance is denoted by $R_{u_j,r}$ and using Al-Kashi formula it is defined as
\begin{equation}
    \begin{split}
        R_{u_j,r}^2 & = R_{u_j}^2+R_r^2-2R_{u_j}R_r\cos(\theta) 
         = (R_r-R_{u_j}\cos(\theta))^2+R_{u_j}^2\sin^2(\theta),
\end{split}
\end{equation}

where:
\begin{itemize}
    \item $\theta$ is the angle between the typical user and the ABS. It is uniformly distributed with PDF $f_{\theta}(x)=\frac{1}{2\pi}\mathbbm{1}_{\left[0,2\pi\right]}(x)$.
    \item $R_{u_j}$ is the horizontal distance between the center of the cluster and the ABS associated with GS located in ring $j$. It is defined as $R_{u_j}=R_j-T_j\cos(\theta_j)$ where $R_j=\frac{2n-1}{2N}R_0$ is the horizontal distance between the GS and the center of the cluster.
    \item $R_r$ is the distance between the typical user and the cluster center. The density probability function of $R_r$ is defined as
    
    \begin{equation}
        f_{R_r}(r)=\left\{\begin{matrix}    \frac{2r}{R_0^2}, & 0\leq r \leq R_0\\ 
        0, & \rm otherwise
                    \end{matrix}.\right.
    \end{equation}
\end{itemize}

The cumulative distribution function (CDF) of $R_{u_j,r}$ is defined as

\begin{equation}
    F_{R_{u_j,r}}(r)= \mathbbm{E}_{\theta}\left( \mathbbm{P}(R_{u_j,r}<r|\theta) \right),
\end{equation}

\begin{equation*}
    \begin{split}
        \text{where} \  \mathbb{P}(R_{u_j,r}<r|\theta) & = \mathbb{P}(R_{u_j,r}^2<r^2|\theta)
         = \mathbb{P}\left((R_r-R_{u_j}\cos(\theta))^2<r^2-R_{u_j}^2\sin^2(\theta)|\theta\right).\\
\end{split}
\end{equation*}

If  $ r^2-R_{u_j}^2\sin^2(\theta)\leq0$, we have $\mathbb{P}(R_{u_j,r}<r|\theta)=0$. Otherwise, we have

\begin{equation}
    \begin{split}
          \mathbb{P}(R_{u_j,r}<r|\theta)=&\mathbb{P}\left(-\sqrt{r^2-R_{u_j}^2\sin^2(\theta)}<R_r-R_{u_j}\cos(\theta)<\sqrt{r^2-R_{u_j}^2\sin^2(\theta)}\ |\theta\right)\\ =&\mathbb{P}\left(R_{u_j}\cos(\theta)-\sqrt{r^2-R_{u_j}^2\sin^2(\theta)}<R_r<R_{u_j}\cos(\theta)+\sqrt{r^2-R_{u_j}^2\sin^2(\theta)} \ |\theta\right)\\  =&F_{R_r}\left(R_{u_j}\cos(\theta)+\sqrt{r^2-R_{u_j}^2\sin^2(\theta)}\right)-F_{R_r}\left(R_{u_j}\cos(\theta)-\sqrt{r^2-R_{u_j}^2\sin^2(\theta)}\right).
    \end{split}
\end{equation}

Thus, to determine the CDF of $R_{u_j,r}$, we need to determine the the range of $\theta$ and the range of $r$ where $0<R_{u_j}\cos(\theta)+\sqrt{r^2-R_{u_j}^2\sin^2(\theta)}<R_0$, $0<R_{u_j}\cos(\theta)-\sqrt{r^2-R_{u_j}^2\sin^2(\theta)}<R_0$ and $g(r,\theta)=r^2-R_{u_j}^2\sin^2(\theta)>0$.

\textbf{i)} First, we will determine the range of $\theta$ where $g(r,\theta)=r^2-R_{u_j}^2\sin^2(\theta)>0$.

We have $r^2-R_{u_j}^2\sin^2(\theta)>0 \Rightarrow -\frac{r}{R_{u_j}}<\sin(\theta)<\frac{r}{R_{u_j}}$.
 There are two cases:
 
 $*$ If $r\geq R_{u_j}$:
\begin{equation}
    \begin{split}
     -\frac{r}{R_{u_j}}< \sin(\theta)<\frac{r}{R_{u_j}} \Rightarrow -1 \leq \sin(\theta) \leq 1
    \Rightarrow 0 \leq \theta \leq 2\pi.
    \end{split}
\end{equation}      
 
$*$ If $r< R_{u_j}$:
\begin{equation}
    -\frac{r}{R_{u_j}}< \sin(\theta)<\frac{r}{R_{u_j}} \Rightarrow
    \left\{\begin{matrix}
    0<\theta<\arcsin(\frac{r}{R_{u_j}})\\ 
    \pi-\arcsin(\frac{r}{R_{u_j}})<\theta<\pi+\arcsin(\frac{r}{R_{u_j}})\\ 
    2\pi-\arcsin(\frac{r}{R_{u_j}})<\theta<2\pi
    \end{matrix}.\right.
\end{equation}
\textbf{ii)} Second, we will determine the range of $\theta$ where $R_{u_j}\cos(\theta)+\sqrt{g(r,\theta)}<R_0 $.

$R_{u_j}\cos(\theta)+\sqrt{g(r,\theta)}<R_0 \Rightarrow \sqrt{g(r,\theta)}<R_0-R_{u_j}\cos(\theta) \Rightarrow g(r,\theta)<\left(R_0-R_{u_j}\cos(\theta)\right)^2 \Rightarrow r^2-R_{u_j}^2\sin^2(\theta)<R_0^2-2R_0R_{u_j}\cos(\theta)+R_{u_j}^2\cos^2(\theta) \Rightarrow \cos(\theta)<\frac{R_0^2+R_{u_j}^2-r^2}{2R_0R_{u_j}}$.

$*$ If $r<R_0-R_{u_j}$, we have $   \frac{R_0^2+R_{u_j}^2-r^2}{2R_0R_{u_j}}>1$. So,
\begin{equation}
    \cos(\theta)<\frac{R_0^2+R_{u_j}^2-r^2}{2R_0R_{u_j}} \Longrightarrow  0<\theta<2\pi.
\end{equation}

$*$ If $R_0-R_{u_j}<r<\sqrt{R_0^2+R_{u_j}^2}$, we have $0<\frac{R_0^2+R_{u_j}^2-r^2}{2R_0R_{u_j}}<1$. So,
\begin{equation}
    \cos(\theta)<\frac{R_0^2+R_{u_j}^2-r^2}{2R_0R_{u_j}} \Longrightarrow  \arccos(\frac{R_0^2+R_{u_j}^2-r^2}{2R_0R_{u_j}})<\theta<2\pi-\arccos(\frac{R_0^2+R_{u_j}^2-r^2}{2R_0R_{u_j}}).
\end{equation}

$*$ If $\sqrt{R_0^2+R_{u_j}^2}<r<R_0+R_{u_j}$, we have $ -1<\frac{R_0^2+R_{u_j}^2-r^2}{2R_0R_{u_j}}<0$. So,
\begin{equation}
    \cos(\theta)<\frac{R_0^2+R_{u_j}^2-r^2}{2R_0R_{u_j}} \Longrightarrow  \pi-\arccos(-\frac{R_0^2+R_{u_j}^2-r^2}{2R_0R_{u_j}})<\theta<\pi+\arccos(-\frac{R_0^2+R_{u_j}^2-r^2}{2R_0R_{u_j}}).
\end{equation}
\textbf{iii)} Third, we will determine the range of $\theta$ where $R_{u_j}\cos(\theta)+\sqrt{g(r,\theta)}>0 $.

$* cos(\theta)>0 \Longrightarrow R_{u_j}\cos(\theta)+\sqrt{g(r,\theta)}>0, \quad \forall r $.

$* cos(\theta)<0 \Longrightarrow R_{u_j}\cos(\theta)+\sqrt{g(r,\theta)}>0 \Rightarrow \sqrt{g(r,\theta)}>-R_{u_j}\cos(\theta) \Rightarrow g(r,\theta)>R_{u_j}^2\cos^2(\theta)  \Rightarrow r^2 - R_{u_j}^2\sin^2(\theta)>R_{u_j}^2\cos^2(\theta) \Rightarrow r^2>R_{u_j}^2 \Rightarrow r>R_{u_j}$.

\begin{equation}
    \Longrightarrow
    \left\{\begin{matrix}
    \text{If } r<R_{u_j},& \quad R_{u_j}\cos(\theta)+\sqrt{g(r,\theta)}>0, \ \forall \theta \in \left [0,\frac{\pi}{2}\right] \bigcup \left [\frac{3\pi}{2},2\pi\right].\\ 
    \text{If } r>R_{u_j},& \quad R_{u_j}\cos(\theta)+\sqrt{g(r,\theta)}>0, \ \forall \theta \in \left [0,2\pi  \right ].
    \end{matrix}\right.
\end{equation}
\textbf{v)} Forth, we will determine the range of $\theta$ where $R_{u_j}\cos(\theta)-\sqrt{g(r,\theta)}>0 $.

$* cos(\theta)>0 \Longrightarrow R_{u_j}\cos(\theta)-\sqrt{g(r,\theta)}>0 \Rightarrow R_{u_j}\cos(\theta)>\sqrt{g(r,\theta)} \Rightarrow R_{u_j}^2\cos^2(\theta)>g(r,\theta)  \Rightarrow R_{u_j}^2\cos^2(\theta)>r^2 - R_{u_j}^2\sin^2(\theta) \Rightarrow R_{u_j}^2>r^2 \Rightarrow r<R_{u_j}$.

$* cos(\theta)<0 \Longrightarrow R_{u_j}\cos(\theta)-\sqrt{g(r,\theta)}<0, \quad \forall r $.

\begin{equation}
    \Longrightarrow
    \left\{\begin{matrix}
    \text{If } r<R_{u_j},& \quad R_{u_j}\cos(\theta)-\sqrt{g(r,\theta)}>0, \ \forall \theta \in \left [0,\frac{\pi}{2}\right] \bigcup \left [\frac{3\pi}{2},2\pi\right].\\ 
    \text{If } r>R_{u_j},& \quad R_{u_j}\cos(\theta)-\sqrt{g(r,\theta)}<0, \ \forall \theta \in \left [0,2\pi  \right ].
    \end{matrix}\right.
\end{equation}
\textbf{iv)} Fifth, we will determine the range of $\theta$ where $R_{u_j}\cos(\theta)-\sqrt{g(r,\theta)}<R_0 $.

$R_{u_j}\cos(\theta)-\sqrt{g(r,\theta)}<R_0 \Rightarrow R_{u_j}\cos(\theta)-R_0<\sqrt{g(r,\theta)}$.

We have, $R_{u_j}\cos(\theta)-R_0<R_{u_j}-R_0<0$. So, 
\begin{equation}
    R_{u_j}\cos(\theta)-\sqrt{g(r,\theta)}<R_0, \ \forall \theta \in \left [0,2\pi  \right ], \ \forall r \geq 0.
\end{equation}

\section{Proof of Lemma~\ref{Lm3}}\label{app:Lm3}
$\mathbb{A}_{\rm N_i}$ is the probability that the average power received at the typical user from the closest NLoS ABS associated with GS located in ring $i$ is greater than the powers received from the other types of BS. It is mathematically defined as 
\begin{equation}\label{A_N}
    \begin{split}
        \mathbb{A}_{\rm N_i}(r)=&\prod_{j \in \left[1,N\right]}  \mathbb{P}(P^r_{\rm N_i}>P^r_{\rm L_j}|D_{\rm N_i}=r) \times \prod_{j \in \left[1,N\right], j \neq i} \mathbb{P}(P^r_{\rm N_i}>P^r_{\rm N_j}|D_{\rm N_i}=r) \\
        &\times\left(\sum_{j\in \left[1,N\right]}p_j\mathbb{P}(P^r_{\rm N_i}>P^r_{\rm CABS_j}|D_{\rm N_i}=r)+p_{\rm out}\right)  \times \mathbb{P}(P^r_{\rm N_i}>P^r_{\rm T}|D_{\rm N_i}=r). 
    \end{split}
\end{equation}

Using the null probabilities of $\Phi_{\rm L_j}$, $\Phi_{\rm N_j}$ and $\Phi_{\rm T}$, we have the following results

\begin{equation}\label{NL}
    \begin{split}
        \mathbb{P}&(P^r_{\rm N_i}>P^r_{\rm L_j}|D_{\rm N_i}=r)=\mathbb{P}\left( \rho_{\rm ABS}\eta_{\rm N}D_{\rm N_i}^{-\alpha_{\rm N}}> \rho_{\rm ABS}\eta_{\rm L}D_{\rm L_j}^{-\alpha_{\rm L}}|D_{\rm N_i}=r\right)\\
        &=\mathbb{P}\left(D_{\rm L_j}> \left( \frac{\eta_{\rm L}}{\eta_{\rm N}} \right)^{\frac{1}{\alpha_{\rm L}}}r^{\frac{\alpha_{\rm N}}{\alpha_{\rm L}}} \right)
        = \exp\left(-2\pi p_j\delta\lambda_C \int_{0}^{\sqrt{d_{\rm NL}^2(r)-h_{u_j}^2}}xp_{L_j}(x){\rm d}x\right).
    \end{split}
\end{equation}
\begin{equation}\label{NN}
    \begin{split}
        \mathbb{P}(P^r_{\rm N_i}>P^r_{\rm N_j}|D_{\rm N_i}=r)=&\mathbb{P}\left( \rho_{\rm ABS}\eta_{\rm N}D_{\rm N_i}^{-\alpha_{\rm N}}> \rho_{\rm ABS}\eta_{\rm N}D_{\rm N_j}^{-\alpha_{\rm N}}|D_{\rm N_i}=r\right)
        = \mathbb{P}(D_{\rm N_j}>r)\\
        = &\exp\left(-2\pi p_j\delta\lambda_C \int_{0}^{\sqrt{\max(r,h_{u_j})^2-h_{u_j}^2}}xp_{N_j}(x){\rm d}x\right).
    \end{split}
\end{equation}
\begin{equation}\label{NT}
    \begin{split}
        \mathbb{P}(P^r_{\rm N_i}>P^r_{\rm T}|D_{\rm N_i}=r)=&\mathbb{P}\left( \rho_{\rm ABS}\eta_{\rm N}D_{\rm N_i}^{-\alpha_{\rm N}}> \rho_{\rm T}\eta_{\rm T}D_{\rm T}^{-\alpha_{\rm T}}|D_{\rm N_i}=r\right)\\
        =\mathbb{P}(D_{\rm T}>& \left( \frac{\rho_{\rm T}\eta_{\rm T}}{\rho_{\rm ABS}\eta_{\rm N}} \right)^{\frac{1}{\alpha_{\rm T}}}r^{\frac{\alpha_{\rm N}}{\alpha_{\rm T}}} )
        = \exp\left(-\pi\lambda_{\rm T}d_{\rm NT}^2(r)\right)\mathbbm{1}_{[0,+\infty[}(r).
    \end{split}
\end{equation}

In addition, we have
\begin{equation}
    \begin{split}
        P^r_{\rm N_i}>P^r_{\rm CABS_j}
        \Rightarrow& \left\{\begin{matrix}
        P^r_{\rm N_i}>\rho_{\rm ABS}\eta_{\rm N}G_{\rm N}D_{u_j,r}^{-\alpha_{\rm N}} ,& \qquad \text{if the ABS is a NLoS ABS}\\ 
        P^r_{\rm N_i}>\rho_{\rm ABS}\eta_{\rm L}G_{\rm L}D_{u_j,r}^{-\alpha_{\rm L}} ,& \qquad \text{if the ABS is a LoS ABS}
        \end{matrix}.\right.\\
        \Rightarrow& \left\{\begin{matrix}
        D_{u_j,r}> D_{\rm N_i} ,& \qquad \text{if the ABS is a NLoS ABS}\\ 
        D_{u_j,r}> \left( \frac{\eta_{\rm L}}{\eta_{\rm N}} \right)^{\frac{1}{\alpha_{\rm L}}}D_{\rm N_i}^{\frac{\alpha_{\rm N}}{\alpha_{\rm L}}} ,& \qquad \text{if the ABS is a LoS ABS}
        \end{matrix}.\right.
    \end{split}
\end{equation}

\begin{equation}\label{NC}
\begin{split}
    \mathbb{P}&(P^r_{\rm N_i}>P^r_{\rm CABS_j}|D_{\rm N_i}=r)=  \mathbb{E}_{D_{u_j,r}}\left[ \mathbbm{1}\left(D_{u_j,r}>d_{\rm NL_j}(r) \right)\times p_{L_j}\left( \sqrt{D_{u_j,r}^2-h_{u_j}^2} \right) \right] \\
    &+ \mathbb{E}_{D_{u_j,r}}\left[ \mathbbm{1}\left(D_{u_j,r}>r \right)\times p_{N_j}\left( \sqrt{D_{u_j,r}^2-h_{u_j}^2} \right) \right]
    = \int_{d_{\rm NL_j}(r)}^{+\infty}p_{L_j}\left( \sqrt{x^2-h_{u_j}^2} \right)f_{D_{u_j,r}}(x)\ {\rm d}x\\
    &+ \int_{\max(h_{u_j},r)}^{+\infty}p_{N_j}\left( \sqrt{x^2-h_{u_j}^2} \right)f_{D_{u_j,r}}(x)\ {\rm d}x.
\end{split}
\end{equation}

Substituting (\ref{NL}), (\ref{NN}), (\ref{NT}) and (\ref{NC}) in (\ref{A_N}), we get the final expression provided in (\ref{A_N1}).

\section{Proof of Lemma~\ref{Lm5}}\label{app:Lm5}

$\mathbb{A}_{\rm CL_j}$ is the probability that the average power received at the typical user from the cluster ABS given that it is a LoS ABS associated with GS located in ring $j$ is greater than the powers received from the other types of BS. It is mathematically defined as

\begin{equation}
    \begin{split}
        \mathbb{A}_{\rm CL_j}(r)=&\prod_{i \in \left[1,N\right]} \left( \mathbb{P}(P^r_{\rm CL_j}>P^r_{\rm N_i}|D_{u_j,r}=r) \times \mathbb{P}(P^r_{\rm CL_j}>P^r_{\rm L_i}|D_{u_j,r}=r) \right)
        \times\mathbb{P}(P^r_{\rm CL_j}>P^r_{\rm T}|D_{u_j,r}=r),  
    \end{split}
\end{equation}
where $P^r_{\rm CL_j}=\rho_{\rm ABS}\eta_{\rm L}G_{\rm L}D_{u_j,r}^{-\alpha_{\rm L}}$ is the received power at the typical user from the cluster ABS given that the cluster ABS is a LoS ABS.

Using the null probabilities of $\Phi_{\rm L_j}$, $\Phi_{\rm N_j}$ and $\Phi_{\rm T}$, we have the following results

\begin{equation}
    \begin{split}
        \mathbb{P}&(P^r_{\rm CL_j}>P^r_{\rm N_i}|D_{u_j,r}=r)
        =\mathbb{P}(\rho_{\rm ABS}\eta_{\rm L}D_{u_j,r}^{-\alpha_{\rm L}}>\rho_{\rm ABS}\eta_{\rm N}D_{\rm N_i}^{-\alpha_{\rm N}}|D_{u_j,r}=r)\\ 
        =& \mathbb{P}\left(D_{\rm N_i}>\left( \frac{\eta_{\rm N}}{\eta_{\rm L}} \right)^{\frac{1}{\alpha_{\rm N}}}r^{\frac{\alpha_{\rm L}}{\alpha_{\rm N}}}\right)
        = \exp\left(-2\pi p_i\delta\lambda_C \int_{0}^{\sqrt{d_{\rm LN_i}^2(r)-h_{u_i}^2}}xp_{N_i}(x){\rm d}x\right).
    \end{split}
\end{equation}

\begin{equation}
    \begin{split}
        \mathbb{P}(P^r_{\rm CL_j}&>P^r_{\rm L_i}|D_{u_j,r}=r)
        =\mathbb{P}(\rho_{\rm ABS}\eta_{\rm L}D_{u_j,r}^{-\alpha_{\rm L}}>\rho_{\rm ABS}\eta_{\rm L}D_{\rm L_i}^{-\alpha_{\rm L}}|D_{u_j,r}=r) 
        = \mathbb{P}\left(D_{\rm L_i}>r\right)\\
        =& \exp\left(-2\pi p_i\delta\lambda_C \int_{0}^{\sqrt{\max(r,h_{u_i})^2-h_{u_i}^2}}xp_{L_i}(x){\rm d}x\right).
    \end{split}
\end{equation}

\begin{equation}
    \begin{split}
        \mathbb{P}(P^r_{\rm CL_j}>P^r_{\rm T}|D_{u_j,r}=r)=& \mathbb{P}(\rho_{\rm ABS}\eta_{\rm L}D_{u_j,r}^{-\alpha_{\rm L}}>\rho_{\rm T}\eta_{\rm T}D_{\rm T}^{-\alpha_{\rm T}}|D_{u_j,r}=r)\\
        =& \mathbb{P}\left(D_{\rm T}> \left( \frac{\rho_{\rm T}\eta_{\rm T}}{\rho_{\rm ABS}\eta_{\rm L}} \right)^{\frac{1}{\alpha_{\rm T}}}r^{\frac{\alpha_{\rm L}}{\alpha_{\rm T}}} \right)
        = \exp\left(-\pi\lambda_{\rm T}d_{\rm LT}^2(r)\right).
    \end{split}
\end{equation}

Thus, we can deduce the final expression (\ref{A_CL}). 

\section{Proof of Lemma~\ref{Lm7}}\label{app:Lm7}
Supposing that the origin of the coordinate system is the typical user and denoting $z_0$ the location of the serving BS, the Laplace transform of the interference power conditioned on the serving BS being at a distance $r$ from the typical user is defined as

\begin{equation}
    \begin{split}
        &L_I(s)= \mathbb{E}_I\left[\exp(-sI) \right]\\
        &\overset{(a)}{=} \mathbb{E}_{\Phi_{\rm T}}\left[\prod_{y\in\Phi_{\rm T}\setminus\left\{z_0\right\}}\mathbb{E}_{G_{\rm T}}\exp(-s P^r_{\rm T,y}) \right]
        \times \prod_{i\in\left[1,N\right]}\mathbb{E}_{\Phi_{\rm L_i}}\left[\prod_{y\in\Phi_{\rm L_i}\setminus\left\{z_0\right\}}\mathbb{E}_{G_{\rm L}}\exp(-s P^r_{\rm L_i,y}) \right]\\
        & \times \prod_{i\in\left[1,N\right]}\mathbb{E}_{\Phi_{\rm N_i}}\left[\prod_{y\in\Phi_{\rm N_i}\setminus\left\{z_0\right\}}\mathbb{E}_{G_{\rm N}}\exp(-s P^r_{\rm N_i,y}) \right]\times \left(p_{\rm out}+ \sum_{j\in\left[1,N\right]} p_j \mathbb{E}_{I}\exp(-s P^r_{\rm CABS_j})\right)\\
        &\overset{(b)}{=} \mathbb{E}_{\Phi_{\rm T}}\left[\prod_{y\in\Phi_{\rm T}\setminus\left\{z_0\right\}} \left(\frac{m_{\rm T}}{m_{\rm T}+s \rho_{\rm T}\eta_{\rm T}D_{\rm T,y}^{-\alpha_{\rm T}}} \right)^{m_{\rm T}} \right]
        \times \prod_{i\in\left[1,N\right]}\mathbb{E}_{\Phi_{\rm L_i}}\left[\prod_{y\in\Phi_{\rm L_i}\setminus\left\{z_0\right\}}\left(\frac{m_{\rm L}}{m_{\rm L}+s \rho_{\rm ABS}\eta_{\rm L}D_{\rm L_i,y}^{-\alpha_{\rm L}}} \right)^{m_{\rm L}} \right]\\
        & \times \prod_{i\in\left[1,N\right]}\mathbb{E}_{\Phi_{\rm N_i}}\left[\prod_{y\in\Phi_{\rm N_i}\setminus\left\{z_0\right\}}\left(\frac{m_{\rm N}}{m_{\rm N}+s \rho_{\rm ABS}\eta_{\rm N}D_{\rm N_i,y}^{-\alpha_{\rm N}}} \right)^{m_{\rm N}} \right]\\
        & \times \left(p_{\rm out}+\sum_{j\in\left[1,N\right]} \frac{p_j}{k_j(r)} 
        \begin{pmatrix}
        \int^{\infty}_{e(r)} \left(\frac{m_{\rm L}}{m_{\rm L}+s \rho_{\rm ABS}\eta_{\rm L}x^{-\alpha_{\rm L}}} \right)^{m_{\rm L}}p_{L_j}(\sqrt{x^2-h_{u_j}^2}) f_{D_{u_j,r}}(x)\ {\rm d}x \\ 
        + \int^{\infty}_{l(r)}\left(\frac{m_{\rm N}}{m_{\rm N}+s \rho_{\rm ABS}\eta_{\rm N}x^{-\alpha_{\rm N}}} \right)^{m_{\rm N}}p_{N_j}(\sqrt{x^2-h_{u_j}^2}) f_{D_{u_j,r}}(x)\ {\rm d}x
        \end{pmatrix}\right).\\
    \end{split}
\end{equation}

Step (a) follows from the (\ref{interference}), and the independence of the spatial point process and small-scale fading. Knowing that $\Gamma(m)=(m-1)! $ , (b) follows from the fact that

\begin{equation}
    \begin{split}
        \mathbb{E}_{G_{\rm T}}\exp(-s P^r_{\rm T})=& \int_0^{\infty} f_{G_{\rm T}}(g)\exp(-s P^r_{\rm T})\ {\rm d}g
        = \left(\frac{m_{\rm T}}{m_{\rm T}+s \rho_{\rm T}\eta_{\rm T}D_{\rm T}^{-\alpha_{\rm T}}} \right)^{m_{\rm T}}.
    \end{split}
\end{equation}

Similarly and by replacing the index $\rm T$ with $\rm L$ and $\rm N$, the expression of $\mathbb{E}_{G_{\rm L}}\exp(-s P^r_{\rm L_i})$ and $\mathbb{E}_{G_{\rm N}}\exp(-s P^r_{\rm N_i})$ are derived as follows

\begin{equation}\label{E_G_L}
    \mathbb{E}_{G_{\rm L}}\exp(-s P^r_{\rm L_i})= \left(\frac{m_{\rm L}}{m_{\rm L}+s \rho_{\rm ABS}\eta_{\rm L}D_{\rm L_i}^{-\alpha_{\rm L}}} \right)^{m_{\rm L}}, \mathbb{E}_{G_{\rm N}}\exp(-s P^r_{\rm N_i})= \left(\frac{m_{\rm N}}{m_{\rm N}+s \rho_{\rm ABS}\eta_{\rm N}D_{\rm N_i}^{-\alpha_{\rm N}}} \right)^{m_{\rm N}}.
\end{equation}

In addition, we have 
\begin{equation}
    \begin{split}
        \mathbb{E}_{I}\exp(-s P^r_{\rm CABS_j})=& \mathbb{E}_{D_{u_j,r}}\left[\mathbb{E}_{G_{\rm L}}\exp(-s P^r_{\rm CL_j})\times p_{L_j}\left( \sqrt{D_{u_j,r}^2-h_{u_j}^2}\right)|P^r_{\rm CABS_j}<P^r_{\rm S}\right]\\
        +&\mathbb{E}_{D_{u_j,r}}\left[\mathbb{E}_{G_{\rm N}}\exp(-s P^r_{\rm CN_j})\times p_{N_j}\left( \sqrt{D_{u_j,r}^2-h_{u_j}^2}\right)|P^r_{\rm CABS_j}<P^r_{\rm S} \right],
    \end{split}
\end{equation}
where $P^r_{\rm CL_j}$ and $P^r_{\rm CN_j}$ are the received power at the typical user from the cluster ABS given that the cluster ABS is a LoS ABS and a NLoS ABS, respectively. $P^r_{\rm S}$ is the received power at the typical user from the serving base station.
Using equation (\ref{E_G_L}), we deduce that 
\begin{equation}
    \begin{split}
        \mathbb{E}_{I}\exp(-s P^r_{\rm CABS_j})=&\int^{\infty}_{e(r)} \left(\frac{m_{\rm L}}{m_{\rm L}+s \rho_{\rm ABS}\eta_{\rm L}x^{-\alpha_{\rm L}}} \right)^{m_{\rm L}}p_{L_j}(\sqrt{x^2-h_{u_j}^2}) \frac{f_{D_{u_j,r}}(x)}{k_j(r)}\ {\rm d}x \\
        &+ \int^{\infty}_{l(r)}\left(\frac{m_{\rm N}}{m_{\rm N}+s \rho_{\rm ABS}\eta_{\rm N}x^{-\alpha_{\rm N}}} \right)^{m_{\rm N}}p_{N_j}(\sqrt{x^2-h_{u_j}^2}) \frac{f_{D_{u_j,r}}(x)}{k_j(r)}\ {\rm d}x,
    \end{split} 
\end{equation}
where $k_j(r)=\mathbb{P}(P^r_{\rm CABS_j}<P^r_{\rm S}|D_s=r)$, $S \in \left \{\rm T, L_i, N_i\right \}$, and $e(r)$ and $l(r)$ are the distances to the nearest interfering BS, defined in Remark\ref{RQ}. The variables $k_j(r)$, $e(r)$ and $l(r)$ depend on the type of the serving BS. If the cluster ABS is the serving BS, we have $\mathbb{E}_{I}\exp(-s P^r_{\rm CABS_j})=1$. 

Finally, using the probability generating functional (PGFL) of the PPPs and the distances to the nearest interfering BS from each type of BS , we deduce the expressions (\ref{L_I}) and (\ref{L_IC}).

\section{Proof of Theorem~\ref{thm2}}\label{app:thm2}



The conditional coverage probability, given that the serving BS is a TBS located at a distance $r$ from the typical user, is defined as
\begin{equation}
    \begin{split}
        P^{\rm T}_{\rm cond}(r)=& \mathbb{P}\left( \frac{\rho_{\rm T}\eta_{\rm T}G_{\rm T}r^{-\alpha_{\rm T}}}{\sigma^2+I}>\gamma \right)
        =\mathbb{P}\left(G_{\rm T}>(\rho_{\rm T}\eta_{\rm T})^{-1} r^{\alpha_{\rm T}} (\sigma^2+I)\gamma \right).
    \end{split}     
\end{equation}

The complementary 
CDF of the Gamma distribution can be written as 
$
    \mathbb{P}(G_{\rm T}>g)=\frac{\Gamma_u(m_{\rm T},m_{\rm T}g)}{\Gamma(m_{\rm T})}= \sum_{k=0}^{m_{\rm T}-1}\frac{(m_{\rm T}g)^k}{k!}\exp(-m_{\rm T}g),
$
where $\Gamma_u$ is the upper incomplete Gamma function.

Denoting $U=\sigma^2+I$, $\mu_{\rm T}(r)=m_{\rm T}\gamma (\rho_{\rm T}\eta_{\rm T})^{-1} r^{\alpha_{\rm T}}$, we have

\begin{equation}
    \begin{split}
        P^{\rm T}_{\rm cond}(r)=&\mathbb{E}_U\left[ \sum_{k=0}^{m_{\rm T}-1}\frac{(\mu_{\rm T}(r)U)^k}{k!}\exp(-\mu_{\rm T}(r)U) \right]
        = \sum_{k=0}^{m_{\rm T}-1}\frac{(\mu_{\rm T}(r))^k}{k!}\mathbb{E}_U\left[\exp(-\mu_{\rm T}(r)U)U^k \right].
    \end{split}
\end{equation}

Since $\mathbb{E}_U\left[\exp(-sU)U^k\right]=(-1)^k\frac{\partial^k}{\partial s^k}L_U(s)$, we get
$
    P^{\rm T}_{\rm cond}(r)=\sum_{k=0}^{m_{\rm T}-1} \left[\frac{(-s)^k}{k!}\frac{\partial^k}{\partial s^k}L_U(s) \right]_{s=\mu_{\rm T}(r)} .
$

{\color{black}
We will just provide the proof of $P^{\rm T}_{\rm cond}$. Following the same steps, we can deduce the other conditional coverage probabilities.

The coverage probability given that the serving BS is a TBS is given by
\begin{equation}
    \begin{split}
        P_{\rm cov}^T=& \mathbb{E}_{D_{\rm T}}\left[ P^{\rm T}_{\rm cond}(D_T) A_{\rm T}(D_T)\right]
        = \int_0^{\infty} P^{\rm T}_{\rm cond}(r) A_{\rm T}(r) f_{D_T}(r) \ {\rm d}r ,
    \end{split}
\end{equation}
where $P^{\rm T}_{\rm cond}(r)$ is the conditional coverage probability given that the serving BS is a TBS located at a distance $r$ from the typical user.

The coverage probability given that the serving BS is an ABS is given by
\begin{equation}
    \begin{split}
        P_{\rm cov}^A&= \sum_{i\in\left[1,N\right]}\mathbb{E}_{D_{\rm L_i}}\left[ P^{\rm L_i}_{\rm cond}(D_{\rm L_i}) A_{\rm L_i}(D_{\rm L_i})\right] + \sum_{i\in\left[1,N\right]} \mathbb{E}_{D_{\rm N_i}}\left[P^{\rm N_i}_{\rm cond}(D_{\rm N_i}) A_{\rm N_i}(D_{\rm N_i})\right]\\
        &= \sum_{i\in\left[1,N\right]} \int_{h_{u_i}}^{\infty} P^{\rm L_i}_{\rm cond}(r) A_{\rm L_i}(r) f_{D_{\rm L_i}}(r) \ {\rm d} r + \sum_{i\in\left[1,N\right]} \int_{h_{u_i}}^{\infty} P^{\rm N_i}_{\rm cond}(r) A_{\rm N_i}(r) f_{D_{\rm N_i}}(r) \ {\rm d} r,
    \end{split}
\end{equation}
where $P^{\{ \rm L_i, N_i \}}_{\rm cond}(r)$ is the conditional coverage probability given that the serving BS is a LoS / NLoS ABS located at a distance $r$ from the typical user and associated to a GS in ring $i$. 

The coverage probability given that the serving BS is the cluster ABS is given by
\begin{equation}
    \begin{split}
        P_{\rm cov}^{\rm CA}&= \sum_{j\in\left[1,N\right]} p_j\mathbb{E}_{D_{u_j,r}}\left[ P^{\rm CL_j}_{\rm cond}(D_{u_j,r}) A_{\rm CL_j}(D_{u_j,r})+P^{\rm CN_j}_{\rm cond}(D_{u_j,r}) A_{\rm CN_j}(D_{u_j,r})\right] \\
        &= \sum_{j\in\left[1,N\right]} p_j \int_{h_{u_j}}^{\infty} P^{\rm CL_j}_{\rm cond}(r) A_{\rm CL_j}(r)+P^{\rm CN_j}_{\rm cond}(r) A_{\rm CN_j}(r) f_{D_{u_j,r}} (r) \ {\rm d}r,
    \end{split}
\end{equation}
where $P^{\{ \rm CL_j, CN_j \}}_{\rm cond}(r)$ is the conditional coverage probability given that the serving BS is a cluster LoS/NLoS ABS located at a distance $r$ from the typical user and associated to a GS in ring $j$.

Combining the aforementioned coverage probabilities, we deduce the overall coverage probability as follow
\begin{equation}
    P_{\rm cov}= P^{A}_{\rm cov}+ P^{\rm CA}_{\rm cov} + P^{T}_{\rm cov}.
\end{equation}

}

\section{Proof of Theorem~\ref{thm3}}\label{app:thm3}

The CDF of the Gamma distribution is given by $F_G(g)=\frac{\Gamma_l(m,mg)}{\Gamma(m)}$, and is bounded as \cite{alzer1997some}

\begin{equation}\label{89}
    \left(1-e^{-\beta_1mg}\right)^m<\frac{\Gamma_l(m,mg)}{\Gamma(m)}<\left(1-e^{-\beta_2mg}\right)^m,
\end{equation}
where $\Gamma_l(m,mg)=\int_0^{mg} t^{m-1} e^{-t} dt$, $\beta_1=
\left\{\begin{matrix}
1, \quad &\text{if } m>1\\ 
(m!)^{\frac{-1}{m}}, \quad &\text{if } m<1
\end{matrix}\right.$, $\beta_2=
\left\{\begin{matrix}
(m!)^{\frac{-1}{m}}, \quad &\text{if } m>1\\ 
1, \quad &\text{if } m<1
\end{matrix}\right.$

It has been demonstrated in \cite{bai2014coverage} that the upper bound in (\ref{89}) offers a good approximation to the CDF of the Gamma distribution. As a result, we approximate the coverage probability using the upper bound. Having (\ref{89}) and $\frac{\Gamma_u(m,mg)}{\Gamma(m)}=1-\frac{\Gamma_l(m,mg)}{\Gamma(m)}$, the conditional coverage probability $P_T^C$ can be rewritten as

\begin{equation}
    \begin{split}
        \Tilde{P}_T^C&=1-\mathbb{E}_U\left[\frac{\Gamma_l(m_T,\mu_{\rm T}(r)U)}{\Gamma(m_T)}\right]
        \overset{(a)}{\approx}1- \mathbb{E}_U\left[\left(1-\exp(-\beta_T\mu_{\rm T}(r)U)\right)^{m_T}\right] \\
        &\overset{(b)}{=}\sum_{k=1}^{m_T}\binom{k}{m_T}(-1)^{k+1}  \mathbb{E}_U\left[\exp(-k\beta_T\mu_{\rm T}(r)U)\right]
        =\sum_{k=1}^{m_T}\binom{k}{m_T}(-1)^{k+1} L_U(k\beta_T\mu_{\rm T}(r)),
    \end{split}
\end{equation}
where $\beta_T=(m_T!)^{\frac{-1}{m_T}}$ and $\mu_{\rm T}(r)$ is defined as before. (a) follows from the upper bound in (\ref{89}). (b) follows
from the binomial theorem, the assumption that $m_T$ is
an integer, and the linearity of the expectation. The remaining conditional coverage probability are derived following the same  steps.

\bibliographystyle{IEEEtran}
\bibliography{reference}
\end{document}